\documentclass[12pt,notitlepage,a4paper]{article}

\pdfoutput=1

\usepackage{a4wide}
\usepackage{epsfig}
\usepackage{color,graphicx}
\usepackage{cite}
\usepackage{amssymb,amsmath}
\usepackage{jheppub}

\newcommand{\be}{\begin{equation}}
\newcommand{\ee}{\end{equation}}
\newcommand{\bea}{\begin{eqnarray}}
\newcommand{\eea}{\end{eqnarray}}

\usepackage[dvipsnames]{xcolor}

\usepackage{graphics, color,tikz,mathtools, colortbl, xcolor,enumerate, tikz-3dplot, slashed,bm}

\usetikzlibrary{positioning}
\usetikzlibrary{chains}
\usetikzlibrary{arrows, arrows.meta ,fit,decorations.pathreplacing,decorations.pathmorphing}
\tikzstyle{every picture}+=[remember picture]
\tikzstyle{na} = [baseline]

\usetikzlibrary{arrows, decorations.markings, calc, fadings, decorations.pathreplacing, patterns, decorations.pathmorphing, positioning}
\tikzset{>={Latex[width=1.5mm,length=1.5mm]}}
\usetikzlibrary{graphs,graphs.standard,quotes}

\tikzset{->-/.style={decoration={
  markings,
  mark=at position #1 with {\arrow{>}}},postaction={decorate}}}    
\tikzset{-<-/.style={decoration={
  markings,
  mark=at position #1 with {\arrow{<}}},postaction={decorate}}}  
  
\newcommand{\nn}{\nonumber}

\def\gd{\delta}

\begin{document}

\begin{center}  

\vskip 2cm 

\centerline{\Large {\bf\boldmath 5d to 3d compactifications and discrete anomalies}}

\vskip 1cm

\renewcommand{\thefootnote}{\fnsymbol{footnote}}

   \centerline{
    {\large \bf Matteo Sacchi${}^{a}$} \footnote{matteo.sacchi@maths.ox.ac.uk}{\large \bf , Orr Sela${}^{b}$} \footnote{osela@physics.ucla.edu} {\large \bf and Gabi Zafrir${}^{c,d,e}$} \footnote{gabi.zafrir@oranim.ac.il}}
      
\vspace{1cm}
\centerline{{\it ${}^a$ Mathematical Institute, University of Oxford, Andrew-Wiles Building, Woodstock Road,}}
\centerline{{\it Oxford, OX2 6GG, United Kingdom}}
\centerline{{\it ${}^b$ Mani L. Bhaumik Institute for Theoretical Physics, Department of Physics and Astronomy,}}
\centerline{{\it University of California, Los Angeles, CA 90095, USA}}
\centerline{{\it ${}^c$ C.~N.~Yang Institute for Theoretical Physics,  Stony Brook University, Stony Brook,}}
\centerline{{\it NY 11794-3840, USA}}
\centerline{{\it ${}^d$ Simons Center for Geometry and Physics, Stony Brook University, Stony Brook,}}
\centerline{{\it NY 11794-3840, USA}}
\centerline{{\it ${}^e$ Department of Mathematics and Physics, University of Haifa at Oranim,}}
\centerline{{\it Kiryat Tivon 3600600, Israel}}
\vspace{1cm}

\end{center}

\vskip 0.3 cm

\setcounter{footnote}{0}
\renewcommand{\thefootnote}{\arabic{footnote}}   
   
\begin{abstract}

Much insight into the dynamics of quantum field theories can be gained by studying the relationship between field theories in different dimensions. An interesting observation is that when two theories are related by dimensional reduction on a compact surface, their 't Hooft anomalies corresponding to continuous symmetries are also related: the anomaly polynomial of the lower-dimensional theory can be obtained by integrating that of the higher-dimensional one on the compact surface. Naturally, this relation only holds if both theories are even dimensional. This raises the question of whether similar relations can also hold for the case of anomalies in discrete symmetries, which might be true even in odd dimensions. The natural generalization to discrete symmetries is that the anomaly theories, associated with the lower and higher dimensional theories, would be related by reduction on the compact surface. We explore this idea for compactifications of 5d superconformal field theories (SCFTs) to 3d on Riemann surfaces with global-symmetry fluxes. In this context, it can be used both as a check for these compactification constructions and for discovering new anomalies in the 5d SCFTs. This opens the way to applying the same idea of dimensional reduction of the anomaly theory to more general types of compactifications.

\end{abstract}
 
 \newpage
 
\tableofcontents

\section{Introduction}
\label{sec:intro}

An anomaly in quantum field theory (QFT) is a failure of a classical symmetry to persist at the quantum level. It is typically encoded in a non-invariance of the partition function under a gauge transformation of the anomalous symmetry. A consistent QFT should have no anomaly for a gauge symmetry, while a global symmetry can be anomalous but this indicates an obstruction to gauging it. Although this might suggest that anomalies are a bug, they are instead actually a feature, since they provide an extremely powerful tool to investigate the dynamics of QFTs. This is because they are invariant along the renormalization group (RG) flow \cite{tHooft:1979rat}, so we can compute them in a regime where this computation is feasible and the result should be the same as in a more difficult to access regime of the theory.

Anomalies are thus very useful to investigate a large variety of phenomena that can characterize the low energy behaviour of a QFT, such as infra-red (IR) dualities and symmetry enhancements. The former refers to the situation in which two different theories in the ultraviolet (UV) flow to the same fixed point at low energies, while the second occurs when the manifest global symmetry in the UV gets enlarged to a bigger group in the IR. The anomalies of IR dual theories are expected to match, while in order for two symmetries to combine to form a larger enhanced group their anomalies should satisfy certain necessary conditions.

In this paper we will be interested in exploiting anomalies to study the dimensional reduction of supersymmetric quantum field theories (SQFTs). The latter has proved to be a very efficient way to organize the many instances of IR dualities and symmetry enhancements that are known, as well as discovering new ones. In particular, we are interested in compactifications of $d$-dimensional superconformal field theories (SCFTs) on Riemann surfaces so to obtain $(d-2)$-dimensional theories. These have been studied for 6d $(1,0)$ theories \cite{Gaiotto:2015usa,Ohmori:2015pua,Ohmori:2015pia,Razamat:2016dpl,Bah:2017gph,Kim:2017toz,Nazzal:2018brc,Kim:2018bpg,Kim:2018lfo,Razamat:2018gro,Razamat:2018gbu,Zafrir:2018hkr,Ohmori:2018ona,Sela:2019nqa,Chen:2019njf,Razamat:2019mdt,Pasquetti:2019hxf,Razamat:2019ukg,Razamat:2020bix,Sabag:2020elc,Baume:2021qho,Hwang:2021xyw,Nazzal:2021tiu,Distler:2022yse,Razamat:2022gpm,Heckman:2022suy,Sabag:2022hyw,Kim:2023qbx,Nazzal:2023bzu,Braun:2023fqa,Giacomelli:2023qyc}, for 5d $\mathcal{N}=1$ theories \cite{Sacchi:2021afk,Sacchi:2021wvg,Sacchi:2023rtp}, for 4d $\mathcal{N}=1$ theories \cite{Kutasov:2013ffl,Kutasov:2014hha,Putrov:2015jpa,Gadde:2015wta,Dedushenko:2017osi,Sacchi:2020pet} and for 3d $\mathcal{N}=2$ theories \cite{Benini:2022bwa}. In this context, anomalies are a powerful tool to test whether the conjectured lower dimensional theory that results from the compactification can be correct. In even dimensions, one can indeed match the continuous anomalies of the $(d-2)$-dimensional theory with those obtained by compactifying the anomaly polynomial of the initial $d$-dimensional theory on the Riemann surface. 

The aim of this paper is to investigate whether an analogous dimensional reduction of anomalies can be done for theories in odd dimensions. In this setup there are no continuous anomalies, but there can be discrete ones. More precisely, in this paper we refer with ``discrete anomalies" to anomalies for finite symmetries such as $\mathbb{Z}_r$ or for continuous symmetries but which involve a characteristic class that is valued in a finite group, such as a Stiefel--Whitney class. Moreover, we only focus on anomalies for ordinary higher-form symmetries \cite{Gaiotto:2014kfa}, while we do not consider models with a higher-group or a non-invertible symmetry. For such discrete anomalies there is no anomaly polynomial, but these can still usually be encoded via anomaly inflow \cite{Callan:1984sa} in an invertible topological theory, called \emph{anomaly theory} or \emph{symmetry topological field theory} \cite{Freed:2014iua,Hsin:2018vcg,Monnier:2019ytc,Gaiotto:2020iye,Apruzzi:2021nmk,Burbano:2021loy,Apruzzi:2022dlm,Freed:2022qnc,Freed:2022iao,Kaidi:2022cpf,Bergman:2022otk,vanBeest:2022fss,Kaidi:2023maf}. For a QFT $\mathcal{T}_d$ living in a $d$-dimensional spacetime $X_d$, the anomaly theory $\mathcal{A}_{d+1}$ is a classical $(d+1)$-dimensional theory on a manifold $M_{d+1}$ with boundary $\partial M_{d+1}=X_d$, such that the variation of its action under a background gauge transformation for the anomalous symmetry precisely compensates the non-invariance of the partition function of $\mathcal{T}_d$ under the same transformation. The question that we address is whether we can derive the anomaly theory of a lower dimensional QFT by compactifying that of the higher dimensional one. 

At first glance, this seems quite reasonable. Consider taking $X_d=X_{d-2}\times \Sigma$, for $\Sigma$ some compact boundless 2d surface. We expect that in the low-energy limit, we should get an effective description in terms of a $d-2$ dimensional QFT system on $X_{d-2}$, which is the dimensional reduction of $\mathcal{T}_d$ on $\Sigma$. Similarly, the associated anomaly theory can be taken to live on $M_{d+1}=M_{d-1}\times \Sigma$, with $\partial M_{d-1}=X_{d-2}$, where here we use the fact that $\Sigma$ is boundless. In the same vein, we now expect that the anomaly theory reduced on $\Sigma$, will give a new TQFT, now living on $M_{d-1}$, that is in turn the anomaly theory for $\mathcal{T}_d$ reduced on $\Sigma$.\footnote{Here it was important that $\Sigma$ is boundless, as otherwise there would be additional contributions coming from the boundaries of $\Sigma$. In the context of compactifications, these boundaries are usually referred to as punctures, and it is known that in the presence of punctures, anomaly matching in continuous symmetries needs to be supplemented by the puncture contributions \cite{Razamat:2022gpm}. We expect a similar thing to hold also for the case of anomalies in discrete symmetries, though we shall not pursue it here.} As the anomaly theory is a TQFT, it should be easier to analyze its dimensional reduction than a fully-fledged QFT system. Moreover, here we shall mostly work with the case where the action of the TQFT in question is just a functional of the background fields for the global symmetries, $\mathcal{A}$, which we can schematically write as $S_{d+1}=\int_{M_{d+1}} f(\mathcal{A})$. In these cases, the dimensional reduction can be done by simply integrating over $\Sigma$ leading us to expect that $\mathcal{L}_{d-1}=\int_{\Sigma} f(\mathcal{A})$, for $\mathcal{L}_{d-1}$ the Lagrangian of the dimensionally reduced anomaly theory.    

We confirm this expectation in a variety of examples of compactifications of 5d $\mathcal{N}=1$ SCFTs on Riemann surfaces, possibly with fluxes for global symmetries, to 3d $\mathcal{N}=2$ theories. Many 3d models arising from this type of compactifications have been studied in \cite{Sacchi:2021afk,Sacchi:2021wvg,Sacchi:2023rtp}. Discrete anomalies in these 3d theories can be easily computed with field theory techniques, see e.g.~\cite{Tachikawa:2019dvq,Bergman:2020ifi,Beratto:2021xmn,Genolini:2022mpi,Bhardwaj:2022dyt,Mekareeya:2022spm,Bhardwaj:2023zix,noppy}, and the superconformal index \cite{Bhattacharya:2008zy,Kim:2009wb,Imamura:2011su,Krattenthaler:2011da,Kapustin:2011jm,Willett:2016adv} allows us to do that systematically. These are then expected to descend from some discrete anomaly in the 5d SCFT, which in some cases is known \cite{Morrison:2020ool,Albertini:2020mdx,BenettiGenolini:2020doj,Apruzzi:2021vcu,Genolini:2022mpi}.

The purpose of this matching is three-fold. First of all, confirming the expectation that also the anomaly theory encoding the discrete anomalies of a higher dimensional theory can be compactified so to get that of the lower dimensional one, in the particular set-up of the 5d to 3d compactifications, can make us confident in extending this to other situations. One can indeed study the dimensional reduction of discrete anomalies under more general compactifications than those on Riemann surfaces. For example, one could compare the discrete anomalies of a 6d SCFT with those of the 3d theory obtained from compactification on a 3-manifold \cite{Dimofte:2011ju,Gang:2018wek}. Hence, this would open the way to investigating these less explored aspects of other known compactifications of QFTs.

Another important application is to confirm the results of \cite{Sacchi:2021afk,Sacchi:2021wvg,Sacchi:2023rtp} for the compactifications of 5d $\mathcal{N}=1$ SCFTs to 3d $\mathcal{N}=2$ theories by matching the dimensional reduction of known 5d anomalies with the anomalies computed directly in 3d. Specifically, we will match
\begin{itemize}
\item the mixed anomaly \cite{BenettiGenolini:2020doj} of the 5d rank 1 $E_1$ Seiberg SCFT \cite{Seiberg:1996bd} between its $\mathbb{Z}_2^{[1]}$ 1-form symmetry \cite{Morrison:2020ool,Albertini:2020mdx} and its $SO(3)$ flavor symmetry \cite{Kim:2012gu,Cremonesi:2015lsa, Apruzzi:2021vcu}, after torus compactification with flux \cite{Sacchi:2021afk};
\item the Witten anomaly \cite{Witten:1982fp} of the rank 1 $E_{N_f+1}$ SCFTs after compactification on a genus $g$ Riemann surface, as well as that of the higher rank generalizations that UV complete some of the 5d SQCD $SU(N+1)$ theories after torus compactification with flux \cite{Sacchi:2021afk,Sacchi:2021wvg}.
\end{itemize}

Finally, one can reverse the logic and use the 3d field theories of \cite{Sacchi:2021afk,Sacchi:2021wvg,Sacchi:2023rtp} to argue for the presence of some discrete anomalies in the 5d SCFTs that are not known. More precisely, we will find that
\begin{itemize}
\item the rank 1 $E_0$ SCFT has a mixed anomaly between its $\mathbb{Z}_3^{[1]}$ 1-form symmetry \cite{Morrison:2020ool,Albertini:2020mdx} and the $SU(2)_R$ R-symmetry of the form
\be
\frac{2\pi i}{3}\int_{M_6} B_2( C_2 (R) \text{ mod } 3)\,,
\ee
where $B_2$ is the background field for the 1-form symmetry and $C_2(R)$ is the second Chern class for the R-symmetry;
\item the rank 1 $E_6$ SCFT has a mixed anomaly between its $E_6/\mathbb{Z}_3$ 0-form flavor symmetry and the R-symmetry of the form
\be
\frac{4\pi i}{3}\int_{M_6} w_2 (E_6/\mathbb{Z}_3) \left(C_2 (R)\text{ mod }3\right) \,,
\ee
where $w_2 (E_6/\mathbb{Z}_3)$ is the second Stiefel--Whitney class of $E_6/\mathbb{Z}_3$;
\item the rank 1 $E_3$ SCFT has a mixed anomaly between its $PSU(3)$ 0-form flavor symmetry and the R-symmetry of the form
\be
\frac{4\pi i}{3}\int_{M_6} w_2 (SU(3)/\mathbb{Z}_3) \left(C_2 (R) \text{ mod }3\right)\,.
\ee
\end{itemize}

The rest of the paper is organized as follows. In Section \ref{sec:mixedanom} we study various mixed discrete anomalies in the 3d models that arise from compactification of 5d SCFTs on Riemann surfaces, possibly with fluxes for the global symmetries. We separate the analysis based on the two types of mixed anomalies that we can have in 3d. In Subsection \ref{sec:anom10form} we study mixed discrete anomalies between a 1-form and a 0-form symmetry in 3d, and then we independently derive them from the compactification of the 5d anomaly theory. The main examples are the compactification of the $E_1$ SCFT on a torus with flux and the compactification of the $E_0$ theory on a genus $g$ surface. In Subsection \ref{sec:anom00form} we instead study mixed discrete anomalies between two 0-form symmetries, one of which is a flavor symmetry and the other the R-symmetry, in 3d and show how these can also be obtained by compactifying a 5d anomaly theory. The main examples are the compactifications of the rank 1 $E_{N_f+1}$ SCFTs on a genus 2 surface with no flux. In Section \ref{sec:wittenanom} we first discuss the Witten anomaly of 5d theories and argue that it should reduce to a parity anomaly for the $U(1)_R$ R-symmetry in 3d. We then match the value of such an anomaly as expected from the 5d picture with the one computed in the 3d models. We focus in particular on the genus $g$ compactifications of the rank 1 $E_{N_f+1}$ SCFTs and the torus compactifications of the higher rank generalizations that UV complete some of the $SU(N+1)$ SQCD theories. We conclude in Section \ref{sec:Disc} with some final considerations.

\section{Mixed discrete anomalies in 3d models from 5d}
\label{sec:mixedanom}

In this section we study various mixed discrete anomalies in some 3d models that arise from compactifications of 5d SCFTs on a Riemann surface, potentially with flux for the global symmetry. For each of the 3d anomalies that we find, we show how the 4d anomaly theory can be derived from compactification of a 6d anomaly theory that encodes some anomaly of the original 5d SCFT. In the cases in which the 5d anomaly is already known we are able to correctly match its compactification with the anomaly computed directly in 3d, while in the other cases our analysis provides new predictions for the anomalies of the 5d SCFTs. We separate our analysis in anomalies between a 1-form and a 0-form symmetry, and between two 0-form symmetries, one of which will be the R-symmetry.

\subsection{Anomalies between 1-form and 0-form symmetries}
\label{sec:anom10form}

In this subsection we study anomalies in the 3d models between a 1-form symmetry and a 0-form symmetry. The 3d theories that can have a 1-form symmetry are those that descend from a 5d SCFT with a 1-form symmetry.\footnote{There can be accidental symmetry enhancements in 3d, but in this case properties of the enhanced symmetry, like anomalies, are not expected to be discernible from 5d. As such, we shall not consider this possibility.} For this reason, we consider two examples: the $E_1$ theory compactified on a torus with flux and the $E_0$ SCFT compactified on a genus $g>1$ surface. The anomaly that we find in the former case can be shown to descend from a 5d anomaly that was discussed in \cite{BenettiGenolini:2020doj}, while the anomaly of the latter theory corresponds to a 5d anomaly that was not known before and so it provides a new prediction for the 5d SCFT.

\subsubsection{$E_1$ SCFT on a torus with flux}
\label{subsec:E1torus}

\subsubsection*{3d analysis}

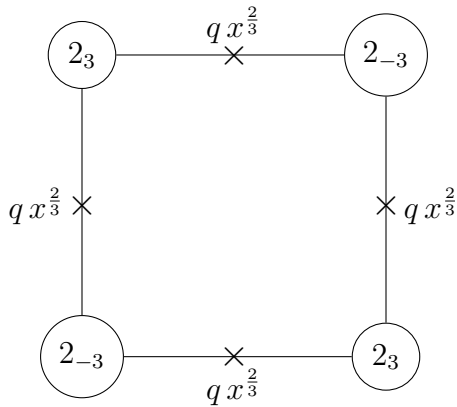
\begin{figure}[t]
\center
\begin{tikzpicture}[baseline=0, font=\scriptsize]
\node[draw, circle] (p1) at (0,0) {\normalsize $2_{3}$};
\node[draw, circle] (p2) at (4,0) {\normalsize $2_{-3}$};
\node[draw, circle] (p3) at (4,-4) {\normalsize $2_{3}$};
\node[draw, circle] (p4) at (0,-4) {\normalsize $2_{-3}$};

\draw[draw, solid] (p1)--(p2)--(p3)--(p4)--(p1);

\node[] at (2,0) {\large $\times$};
\node[] at (4,-2) {\large $\times$};
\node[] at (0,-2) {\large $\times$};
\node[] at (2,-4) {\large $\times$};

\node[] at (2,0.4) {\normalsize $q\,x^{\frac{2}{3}}$};
\node[] at (2,-4.4) {\normalsize $q\,x^{\frac{2}{3}}$};
\node[] at (4.6,-2) {\normalsize $q\,x^{\frac{2}{3}}$};
\node[] at (-0.6,-2) {\normalsize $q\,x^{\frac{2}{3}}$};

\end{tikzpicture}
\caption{The 3d model corresponding to the compactification of the 5d rank 1 $E_1$ SCFT on a torus with flux $F=2$.  In accordance with standard notation, we use circles to denote gauge groups and lines between them denote bifundamental chiral fields. Each cross instead denotes a gauge singlet chiral field that flips the meson constructed with the corresponding bifundamental. The gauge groups are all of type $SU(N)$, with $N$ indicated by the number in the circles and the subscript to this number denoting the level of the Chern--Simons (CS) term if present. Additionally, the power of the fugacity $x$ close to each line indicates the R-charge of the corresponding chiral field under a trial $U(1)_R$ R-symmetry, while the power of $q$ denotes the charge under the $U(1)_q$ symmetry.}
\label{torusNf0}
\end{figure}  

We begin by considering the 3d model coming from the compactification of a 5d SCFT with a 1-form symmetry. Specifically, we consider the 5d rank 1 $E_1$ SCFT on a torus with flux $F$ for the $U(1)$ Cartan of its $SO(3)$ flavor symmetry in a quantization in which the minimal allowed flux is 1. The resulting 3d theory has been studied in \cite{Sacchi:2021afk} and consists of a circular quiver with $2F$ $SU(2)$ gauge nodes with Chern--Simons (CS) level $\pm3$ of alternating signs, where pairs of adjacent nodes are connected by a bifundamental chiral, see Figure \ref{torusNf0} for a depiction of the case $F=2$. There are also $2F$ chiral singlet fields, each of which flips one quadratic invariant built from the bifundamental chirals, which we denote with crosses in the figure. 

The superpotential, on top of the flipping terms, consists only in the monopoles that have minimal integer flux under any pair of adjacent $SU(2)$ gauge nodes simultaneously, dressed with 3 copies of the corresponding bifundamental field so as to make a gauge invariant due to the CS level. This monopole superpotential breaks the $U(1)$ symmetries that rotate independently each bifundamental chiral to their diagonal combination which we denote by $U(1)_q$ and we parameterize as specified in the figure. The parameterization of the R-symmetry that we are using is neither the Cartan of the 5d R-symmetry nor the superconformal one, but it is related to the former by the mixing with $U(1)_q$ given at the level of fugacities by $q\to q\,x^{\frac{1}{3}}$. 

The index of this model for $F=1$ reads (see for example Appendix B of \cite{Sacchi:2021afk} for our index conventions)
\bea
\label{ind_torusNf0}
\mathcal{I}&=&1+\frac{2}{q^2}x^{\frac{2}{3}}+\left(8q^2+\frac{3}{q^4}\right)x^{\frac{4}{3}}+\frac{4}{q^6}x^2+\cdots\,.
\eea
Notice in particular that we have only states with even charges under $U(1)_q$. 

This model also has a $\mathbb{Z}_2^{[1]}$ 1-form symmetry, which comes from the diagonal combination of the $\mathbb{Z}_2$ center symmetries of each $SU(2)$ gauge node that is not screened by the bifundamental matter fields. This is compatible with the fact that the 5d $E_1$ theory also has a $\mathbb{Z}_2^{[1]}$ 1-form symmetry \cite{Morrison:2020ool,Albertini:2020mdx}. We next want to study anomalies involving this symmetry in the 3d model. This can be achieved by considering the variant of the theory in which it is gauged, which has the gauge group $SU(2)^{2F}/\mathbb{Z}_2$. In this theory, there are additional monopole operators corresponding to half-integer fluxes under all gauge groups. These might in general carry global symmetry charges which correspond to a different quantization of these symmetries compared to the theory before gauging (that is, the one with gauge group $SU(2)^{2F}$), implying a mixed anomaly with the 1-form symmetry \cite{Tachikawa:2019dvq,Bergman:2020ifi,Beratto:2021xmn,Genolini:2022mpi,Bhardwaj:2022dyt,Mekareeya:2022spm,Bhardwaj:2023zix}. 

In order to examine these new monopoles, we compute the index of the theory with gauge group $SU(2)^{2F}/\mathbb{Z}_2$, paying special attention to their contribution. For $F=1$ we find
\bea
\label{ind_torusNf0mod2}
\mathcal{I}&=&1+\left(\frac{2}{q^2}+4\zeta q\right)x^{\frac{2}{3}}+\left(8q^2+\frac{3}{q^4}+\frac{4\zeta}{q}\right)x^{\frac{4}{3}}+\left(\frac{4}{q^6}+12\zeta q^3+\frac{4\zeta}{q^3}\right)x^2+\cdots\,,\nn\\
\eea
where $\zeta$ such that $\zeta^2=1$ is the fugacity for the discrete $\mathbb{Z}_2^{[0]}$ magnetic 0-form symmetry that is dual to the 1-form symmetry that we gauged. In particular, the states carrying a non-trivial power of $\zeta$ correspond to the extra monopole operators with half-integer flux for all the gauge groups that are introduced upon gauging $\mathbb{Z}_2^{[1]}$.

Comparing \eqref{ind_torusNf0mod2} and \eqref{ind_torusNf0} we can see that gauging the $\mathbb{Z}_2^{[1]}$ 1-form symmetry has changed the quantization of the charges under $U(1)_q$ from even integers to any integer. This indicates that the theory has a discrete mixed anomaly between these two symmetries of the form
\be
 \pi i\int_{M_4} B_2 (C_1 (U(1)_c)\text{ mod }2)\,,
\ee
where $B_2\in H^2(M_3,\mathbb{Z}_2)$ is the background gauge field for the 1-form symmetry and $C_1 (U(1)_c)$ is the first Chern class of $U(1)_c$, with $c=q^2$.

It is interesting to consider how this result gets generalized for higher values of the flux $F$. We find that the monopole with flux $\frac{1}{2}$ for each $SU(2)$ gauge node contributes to the index as
\be\label{ind_torusNf0mod2genF}
\mathcal{I}\supset 4\zeta q^Fx^{\frac{2F}{3}}\,.
\ee
This indicates the anomaly
\be\label{eq:anomaE1torusF}
\pi i\,F\int_{M_4} B_2 (C_1 (U(1)_c)\text{ mod }2)\,.
\ee
Notice in particular that this is only present for odd $F$, while it vanishes for even $F$.

The result \eqref{ind_torusNf0mod2genF} can be checked by direct computation of the index for low $F$ or by explicitly constructing the corresponding gauge invariant monopole operators and looking at their quantum numbers. Let us first examine the contribution to the index of the bare monopole $T$ with flux $\frac{1}{2}$ for all the gauge nodes. Denoting by $s_i$ the fugacity of the $i$-th $SU(2)$ gauge group (where $i=0,\ldots2F-1$ and $s_{2F}\equiv s_{0}$), this contribution is given by 
\be\label{bareTgenF}
\mathcal{I}\left[T_{\left(\left(\frac{1}{2}\right)^{2F}\right)}\right]=\zeta x^{-\frac{4F}{3}}q^{-2F}g\prod_{i=0}^{F-1}s_{2i}^{3}s_{2i+1}^{-3}\,,
\ee
where the product over gauge fugacities results from the CS terms with ranks of alternating signs ($\pm3$), and the other factors originate from the fermion zero modes of the vector and chiral multiplets. We see that the bare monopole $T$ is not gauge invariant due to the CS terms, and in order to construct a gauge invariant operator $M$ we need to dress it with matter fields. By examining the contribution to the index of the various chiral multiplets in this monopole background, one can check that the monopole $M$ with the lowest R-charge (and therefore with the leading contribution to the index in this magnetic flux sector) is obtained by dressing $T$ with certain components of the bifundamental chirals. More explicitly, denoting by $\chi_{-}^{\left(2i,2i+1\right)}$ the scalar field of the bifundamental chiral with charge $-1$ under the Cartan of the $SU(2)_{2i}$ gauge group and charge $+1$ under the Cartan of $SU(2)_{2i+1}$, its contribution to the index is given by 
\be\label{chimp}
\mathcal{I}\left[\chi_{-}^{\left(2i,2i+1\right)}\right]=q\frac{s_{2i+1}}{s_{2i}}x^{\frac{2}{3}}\,.
\ee
Similarly, denoting by $\chi_{+}^{\left(2i+1,2i+2\right)}$ the corresponding scalar field with charge $+1$ under the Cartan of $SU(2)_{2i+1}$ and charge $-1$ under the Cartan of $SU(2)_{2i+2}$, its contribution is given by 
\be\label{chipm2}
\mathcal{I}\left[\chi_{+}^{\left(2i+1,2i+2\right)}\right]=q\frac{s_{2i+1}}{s_{2i+2}}x^{\frac{2}{3}}\,.
\ee
The gauge invariant operator $M$ that would have the minimal possible R-charge is obtained by dressing $T$ with copies of these two scalar field components so as to cancel the dependence on the gauge fugacities $s_i$ in \eqref{bareTgenF}. There are overall four possible such dressings, given by
\begin{equation*}
M_{\left(\left(\frac{1}{2}\right)^{2F}\right)}^{\left(1\right)}=T_{\left(\left(\frac{1}{2}\right)^{2F}\right)}\prod_{i=0}^{F-1}\left(\chi_{\mp}^{\left(2i,2i+1\right)}\right)^{3}\;,\;M_{\left(\left(\frac{1}{2}\right)^{2F}\right)}^{\left(2\right)}=T_{\left(\left(\frac{1}{2}\right)^{2F}\right)}\prod_{i=0}^{F-1}\left(\chi_{\mp}^{\left(2i,2i+1\right)}\right)^{2}\left(\chi_{\pm}^{\left(2i+1,2i+2\right)}\right)^{1},
\end{equation*}
\begin{equation}
M_{\left(\left(\frac{1}{2}\right)^{2F}\right)}^{\left(3\right)}=T_{\left(\left(\frac{1}{2}\right)^{2F}\right)}\prod_{i=0}^{F-1}\left(\chi_{\mp}^{\left(2i,2i+1\right)}\right)^{1}\left(\chi_{\pm}^{\left(2i+1,2i+2\right)}\right)^{2}\;,\;M_{\left(\left(\frac{1}{2}\right)^{2F}\right)}^{\left(4\right)}=T_{\left(\left(\frac{1}{2}\right)^{2F}\right)}\prod_{i=0}^{F-1}\left(\chi_{\pm}^{\left(2i+1,2i+2\right)}\right)^{3}
\end{equation}
and since they all contribute to the index as 
\be\label{ContM}
\mathcal{I}\left[M_{\left(\left(\frac{1}{2}\right)^{2F}\right)}^{\left(a\right)}\right]=\zeta q^{F}x^{\frac{2F}{3}}\,,\quad\left(a=1,\ldots,4\right)
\ee
we find that the total contribution is given by \eqref{ind_torusNf0mod2genF}, as expected. 

\subsubsection*{Comparison with 5d}

We would like to argue for this anomaly using the 5d construction. The starting point is the symmetries and anomalies of the 5d SCFT. As we already mentioned, the $E_1$ theory has, in addition to the superconformal symmetry, an $SO(3)$ global 0-form symmetry \cite{Kim:2012gu,Cremonesi:2015lsa, Apruzzi:2021vcu} and a $\mathbb{Z}_2^{[1]}$ 1-form symmetry \cite{Morrison:2020ool,Albertini:2020mdx}. It was argued in \cite{BenettiGenolini:2020doj,Apruzzi:2021nmk} that there should be a mixed anomaly between these of the form\footnote{It was further argued in \cite{Apruzzi:2021vcu} that the 1-form and $SO(3)$ symmetries do not form a direct product but rather a 2-group, that is a non-split extension. As such, the anomaly should lift to some anomaly of the full 2-group. However, the extension turns out to trivialize if we break $SO(3)$ down to its Cartan, and as here we shall only consider the case with flux that induces such a background, we shall ignore the proposed 2-group structure.}
\be \label{AnomalyTerm5d}
\frac{\pi i}{2}\int_{M_6}  \mathcal{P}(B_2) w_2 (SO(3)) ,
\ee
where $\mathcal{P}(B_2)$ is the Pontryagin square\footnote{For an ordinary differential 2-form $w$, whose coefficient system is the real numbers, the square would be $w \wedge w$. When the coefficient system becomes discrete though, we can define different squares depending on the coefficent system of the product. The cup product can be used to define a square that preserves the coefficient system, such that if $w \in H^2(M,\mathbb{Z}_2)$ then $w \cup w \in H^4(M,\mathbb{Z}_2)$. However, for the Pontryagin square we instead have that $\mathcal{P}(w) \in H^4(M,\mathbb{Z}_4)$.} of $B_2$, and $w_2 (SO(N))$ denotes the second Stiefel--Whitney class of an $SO(N)$ bundle which is valued mod 2.

Next, we want to consider the compactification of the 5d SCFT on a torus with flux $F$ in the $U(1)$ Cartan of the $SO(3)$ global symmetry. We expect that integrating the anomaly \eqref{AnomalyTerm5d} of the 5d theory on the Riemann surface would yield the anomaly \eqref{eq:anomaE1torusF} of the 3d model.\footnote{Here we shall only consider anomalies involving the $U(1)$ Cartan of the $SO(3)$ symmetry and the 1-form symmetry. In that case, we can also envision having the anomaly terms $\int (B_2)^3$, $\int B_2 w^2_2 (SO(3))$ and $\int B_2 (C_2 (SO(3))\text{ mod 2}))$, where here we are being only schematic. The first one is known to appear in certain $5d$ SCFTs, see \cite{Gukov:2020btk,Apruzzi:2021nmk}, while we are not aware of a $5d$ SCFT realizing the latter two. Either way, for the case at hand only the term in \eqref{AnomalyTerm5d} is known to be present so we need not consider the other terms.} For this we take $M_6 = M_4 \times T^2$ and separate the forms depending on whether they have support on $M_4$ or $T^2$
\bea \label{FormDecomp}
& & w_2 (SO(3)) \rightarrow C_1 (U(1)_c)\text{ mod }2\,,\quad  C_1 (U(1)_c) = C_1^{T^2} (U(1)_c) + C_1^{4d} (U(1)_c) \,, \nonumber\\ 
& & B_2 = B_2^{T^2} + B_2^{S^1} + B_2^{\hat{S}^1} + B_2^{4d}.
\eea
Here in the first line we used the fact that the second Stiefel--Whitney class of $SO(3)$ reduces to the mod 2 first Chern class of the $U(1)_c$ Cartan under the breaking $SO(3) \rightarrow U(1)_c$ such that ${\bf 3} \rightarrow c+1+\frac{1}{c}$. We also use a superscript to denote where the form has support, with $T^2$ signifying support on the torus, $4d$ support on $M_4$ and $S^1$, $\hat{S}^1$ signifying support on one direction in $M_4$ and one of the two circles in $T^2$ respectively. We next consider the forms for which the integral over $T^2$ does not vanish
\be
\int_{T^2} C^{T^2}_1 (U(1)_c)\text{ mod }2 = 2F \text{ mod 2}\,,\quad \int B_2^{T^2} = b\,,\quad \int B_2^{S^1} = A_1^{(1)} , \int B_2^{\hat{S}^1} = A_1^{(2)}\,.
\ee
Here the first term is due to the flux in $U(1)_c$ on $T^2$.\footnote{Previously we were working in terms of $U(1)_q$, which is the Cartan of $\mathfrak{su}(2)$ normalized such that ${\bf 2}\to q+q^{-1}$. This leads to the flux through $T^2$ in $U(1)_q$ units being integer. Instead, in terms of $U(1)_c$ defined such that $c=q^2$ we can see that the minimal charge is $\frac{1}{2}$ and so the flux through $T^2$ in $U(1)_c$ units is an even integer.} The second term takes into account the possibility of introducing an holonomy $b$ in the 1-form symmetry on $T^2$. Finally, the 1-form symmetry potentially leads to two 0-form symmetries when reduced on the two cycles of $T^2$, for which we introduce the background flat connections $A_1^{(1)}$, $A_1^{(2)}\in H^1(M_3,\mathbb{Z}_2)$, and the last two terms provide their relation to the background gauge field for the 1-form symmetry.\footnote{In the language of topological operators, we can consider the three-dimensional topological operator that generates the 1-form symmetry in 5d to wrap the entire Riemann surface giving the topological operator for a 1-form symmetry in 3d, or to wrap only one cycle giving the topological operator for a 0-form symmetry in 3d, or to live entirely in the three-dimensional space and be at a point in the Riemann surface implementing the holonomy.} We can next insert \eqref{FormDecomp} into \eqref{AnomalyTerm5d} and simplify using
\be\label{eq:splitPontryagin}
\mathcal{P}(A+B) = \mathcal{P}(A) + \mathcal{P}(B) + 2 A B \,, 
\ee
ending up with
\bea
& & \frac{\pi i}{2}\int_{T^2 \times M_4} ( C_1^{T^2} (U(1)_c)\text{ mod 2}) \mathcal{P} (B_2^{4d})  + 2(C_1^{4d} (U(1)_c)\text{ mod }2) ( B_2^{T^2} B_2^{4d} + B_2^{S^1}B_2^{\hat{S}^1} ) \nonumber\\  
& = & \pi i \int_{M_4} F \mathcal{P} (B_2^{4d}) + b( C^{4d}_1 (U(1)_c)\text{ mod }2) B_2^{4d} + A_1^{(1)} A_1^{(2)} (C_1^{4d} (U(1)_c)\text{ mod 2})\,.
\eea

Some comments about this result are in order. First, since $\int_{M_4} \mathcal{P} (B_2)$ is even for any spin manifold $M_4$, the anomaly term $F \pi i \int_{M_4}\mathcal{P} (B_2^{4d})$ is non-trivial only for fractional $F$ and signals that the 1-form symmetry cannot be gauged in that case. Instead for $F$ integer, which is the case we are dealing with, this term vanishes which is consistent with our findings. The second term is the one we are mainly interested in, since it gives an anomaly of the form $\pi i \int_{M_4}(C_1^{4d} (U(1)_c)\text{ mod }2) B_2^{4d}$ similar to the one we found in the 3d model \eqref{eq:anomaE1torusF}, when there is a non-trivial holonomy $b$ for the 1-form symmetry on the torus. Finally, the last term is a mixed anomaly between $U(1)_c$ and the two 0-form symmetries descending from the 5d 1-form symmetry. This should be unphysical in the presence of the second anomaly as it can be canceled by redefining $B_2^{4d}\rightarrow B_2^{4d} + A_1^{(1)} A_1^{(2)}$, though it might be physical in cases with $b=0$.

As such, we see that we can naturally understand the 3d field theory results if we assume that there is also an holonomy in the 1-form symmetry on the torus which is $b=F$ mod 2. The question now is how to see that in our compactification set-up there should be such an holonomy. 

For this it is convenient to consider the effect of such an holonomy, where for simplicity we shall consider the 5d gauge theory first before taking the SCFT limit. Consider the basic objects charged under the 1-form symmetry, which are the fundamental Wilson lines. Consider wrapping one around a cycle of the torus and transporting it around the other cycle so that it comes back to itself. In the presence of an holonomy, under such a motion the Wilson line will not come back to itself, but rather to itself acted on by the 1-form symmetry, which in our case means it comes to minus itself.

Next, consider the $N_f=1$ case corresponding to the $E_2$ theory, which can be mass deformed to the $E_1$ theory we are interested in by integrating out the flavor.\footnote{The flavor symmetry algebra of the $E_2$ theory is $\mathfrak{su}(2)\oplus \mathfrak{u}(1)_\beta$ and the mass deformation to $E_1$ is for the $\mathfrak{u}(1)_\beta$ part, thus preserving the $\mathfrak{su}(2)$.} Here we also have fundamental Wilson lines, which will become the fundamental Wilson lines of the $N_f=0$ theory. However, we also have local operators in the fundamental representations, the ones creating and annihilating the flavor quanta. The Wilson line can now end on them and so there is no 1-form symmetry. An alternative way to phrase this is that we can interpret the Wilson line as the world line of the fundamental flavor, and as such the symmetries acting on it should be interpreted as just part of the 0-form $U(1)_F$ flavor symmetry acting on the flavors.

Now consider the fundamental Wilson line wrapped on a cycle of the torus. We can consider the same motion in the $N_f=1$ case as we did in the $N_f=0$ case, where we now interpret this as moving the worldline of the flavor particles such that they sweep the full torus. Now we note the following. The flavors are electrically charged with charge $\pm 1$ under $\mathfrak{u}(1)_F$, and we have magnetic flux in the Cartan of the $\mathfrak{su}(2)$ flavor symmetry of the 5d SCFT. As such we are sweeping an electrically charged object in the background of a magnetic monopole and so it would come back acted by a phase of $e^{2\pi i q_e \mathcal{F}}$, where $q_e$ is the electric charge and $\mathcal{F}$ the flux in some normalization. In order to calculate this phase, we need to consider the relation between $\mathfrak{u}(1)_F$ and the $\mathfrak{su}(2)$ which tells us the correct normalization. It is possible to show that the $\mathfrak{su}(2)\oplus \mathfrak{u}(1)_\beta$ symmetry of the SCFT is related to the $\mathfrak{u}(1)_I \oplus \mathfrak{u}(1)_F$ symmetry of the gauge theory, where $\mathfrak{u}(1)_I$ is the instantonic symmetry, by \cite{Kim:2012gu}
\be
f^2 = \frac{q}{\beta}\,\quad u^4 = q^7 \beta\,,
\ee
where we use the fugacity $q$ for the Cartan of the $\mathfrak{su}(2)$ such that ${\bf 2}\to q+q^{-1}$, $\beta$ for the $\mathfrak{u}(1)_\beta$, $f$ for $\mathfrak{u}(1)_F$ and $u$ for $\mathfrak{u}(1)_I$. This suggests that an operator with charge $1$ under $\mathfrak{u}(1)_F$ carries charge $\frac{1}{2}$ under the Cartan of the $\mathfrak{su}(2)$. This implies that we have $q_e = \frac{1}{2}$, $\mathcal{F} = F$ and so we get the phase $e^{\pi i F}$. In particular, when $F=1$ we have a phase of $-1$. This phase should remain also after we integrate the flavors out, leading to $b= F$ mod $2$, as expected.

\subsubsection{$E_0$ SCFT on a genus $g$ surface}
\label{subsec:E0g}

\subsubsection*{3d analysis}

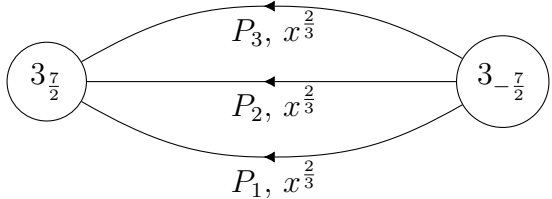
\begin{figure}[t]
\center
\begin{tikzpicture}[baseline=0, font=\scriptsize]
\node[draw, circle] (l) at (0,0) {\normalsize $3_{\frac{7}{2}}$};
\node[draw, circle] (r) at (6,0) {\normalsize $3_{-\frac{7}{2}}$};

\draw[draw, solid,-<] (l)--(3,0);
\draw[draw, solid,-] (3,0)--(r);
\draw[draw, solid,-<] (l) edge [out=30,in=180,loop,looseness=1] (3,1);
\draw[draw, solid,-] (3,1) edge [out=0,in=150,loop,looseness=1] (r);
\draw[draw, solid,-<] (l) edge [out=-30,in=180,loop,looseness=1] (3,-1);
\draw[draw, solid,-] (3,-1) edge [out=0,in=-150,loop,looseness=1] (r);

\node[] at (3,-0.3) {\normalsize $P_2,\,x^{\frac{2}{3}}$};
\node[] at (3,1-0.3) {\normalsize $P_3,\,x^{\frac{2}{3}}$};
\node[] at (3,-1-0.3) {\normalsize $P_1,\,x^{\frac{2}{3}}$};
\end{tikzpicture}
\caption{The 3d model corresponding to the compactification of the 5d rank 1 $E_0$ SCFT on a Riemann surface of genus 2.}
\label{fig:g2E0}
\end{figure}

We now consider the compactification of the 5d rank 1 $E_0$ SCFT on a genus 2 surface. A 3d Lagrangian for this model was proposed in \cite{Sacchi:2023rtp} and we summarize it in Figure \ref{fig:g2E0}, where the superpotential is
\be\label{eq:superpotg2E0}
\mathcal{W}=\sum_{i,j=1}^3P_i(P_j)^2\,.
\ee
In the figure, the power of the fugacity $x$ close to each line indicates the R-charge of the corresponding chiral field under the $U(1)_R$ R-symmetry that descends from the Cartan of the 5d $SU(2)_R$ R-symmetry. This is also the superconformal R-symmetry in this case since there is no abelian global symmetry it can mix with.

This theory has no continuous 0-form symmetry, but it has a $\mathbb{Z}_3^{[1]}$ 1-form symmetry which comes from the diagonal combination of the $\mathbb{Z}_3$ center of each $SU(3)$ gauge node that is not screened by the bifundamental matter fields. This is expected to descend from the $\mathbb{Z}_3^{[1]}$ 1-form symmetry of the 5d $E_0$ theory \cite{Morrison:2020ool,Albertini:2020mdx}.\footnote{The $E_0$ theory also has a $\mathbb{Z}_3$ 0-form symmetry, though we shall not consider it here.} Our goal is to compute anomalies for this symmetry in the 3d theory. 

For this purpose, we compute the index of the theory obtained by gauging $\mathbb{Z}_3^{[1]}$, which corresponds to the variant of the theory with gauge group $[SU(3)\times SU(3)]/\mathbb{Z}_3$. Compared to the theory with gauge group $SU(3)\times SU(3)$, this theory has additional monopole operators corresponding to fluxes that are multiples of $\frac{1}{3}$ for both gauge groups. These might carry charges under other global symmetries that have a different quantization with respect to those with integer fluxes, thus indicating a mixed anomaly with the 1-form symmetry. 

We indeed find that the monopoles with fluxes $(-\frac{1}{3},-\frac{1}{3})$ under both groups (and their Weyl equivalent) give the first non-trivial contribution to the index, which is
\be\label{eq:indg2E0mod3}
9\zeta^2x^{\frac{2}{3}}\,.
\ee
Here $\zeta$ such that $\zeta^3=1$ is the fugacity for the discrete $\mathbb{Z}^{[0]}_3$ magnetic 0-form symmetry that is dual to the 1-form symmetry that we gauged. 

This result should be compared with the one for the variant of the theory with $SU(3)\times SU(3)$ gauge group, which to low order is (see also eq.~(4.27) of \cite{Sacchi:2023rtp})
\be\label{eq:indg2E0}
\mathcal{I}=1+84x^2+327x^4+\cdots\,.
\ee
Comparing \eqref{eq:indg2E0mod3} and \eqref{eq:indg2E0} we can see that while in the theory where $\mathbb{Z}_3^{[1]}$ is not gauged the R-charges come in even integers, in the one where $\mathbb{Z}_3^{[1]}$ is gauged they are quantized as integer multiples of $\frac{2}{3}$. This indicates that there is a mixed anomaly between $\mathbb{Z}_3^{[1]}$ and the R-symmetry which takes the following form:
\be\label{eq:anomE0g2}
\frac{2\pi i}{3}\int_{M_4} B_2 (C_1 (R)\text{ mod } 3)\,,
\ee 
where $B_2\in H^2(M_4,\mathbb{Z}_3)$ is the background field for the 1-form symmetry and $C_1 (R)$ is the first Chern class of the $U(1)_R$ R-symmetry. We point out that this is also the Cartan of the 5d $SU(2)_R$ R-symmetry.

\begin{figure}[t]
	\center
	\includegraphics[width=0.6\textwidth]{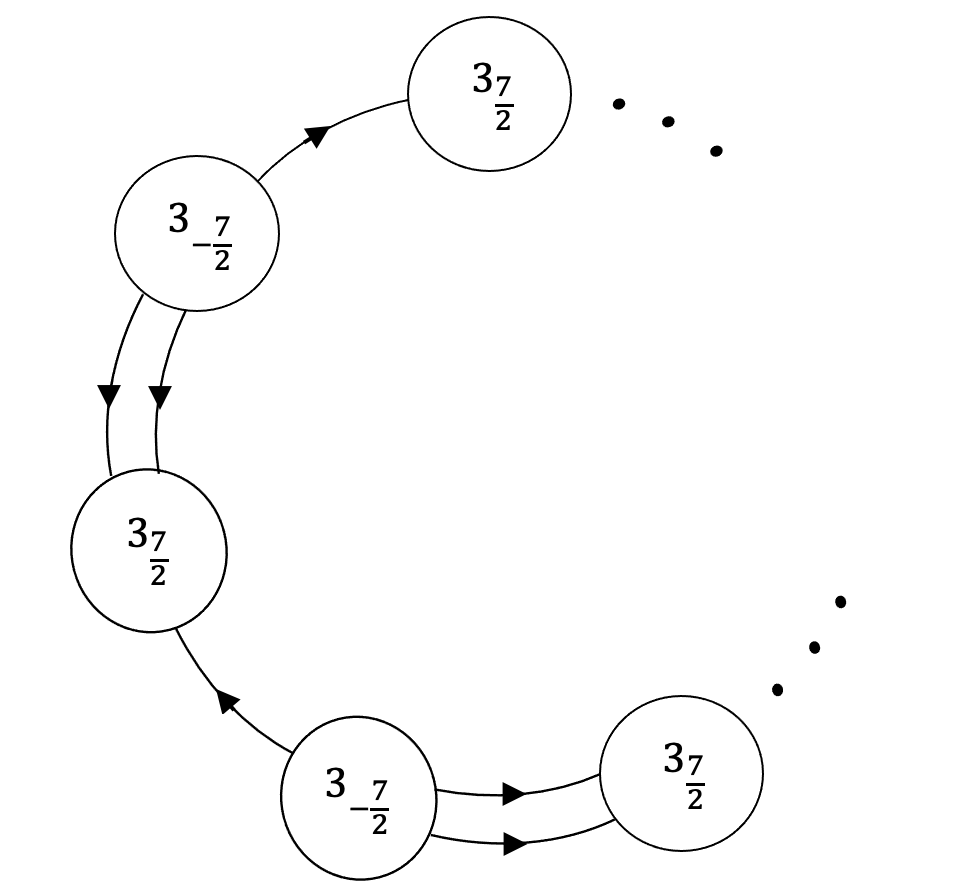} 
	\caption{The 3d theory corresponding to the compactification of the 5d rank-1 $E_0$ SCFT on a genus $g$ surface. The R-charges of all the fields are 2/3 and there is a cubic superpotential given by all the possible cubic invariants of the bifundamental fields.}
	\label{E0_for_g>1}
\end{figure} 

This result can be generalized to the case of the compactification of the 5d $E_0$ SCFT on a Riemann surface of arbitrary genus $g$. In this case, the 3d Lagrangian proposed in \cite{Sacchi:2023rtp} is a generalization of the one depicted in Figure \ref{fig:g2E0} where we have $2(g-1)$ gauge nodes. This is depicted in Figure \ref{E0_for_g>1} and again there is a cubic superpotential given by all the possible cubic invariants of the bifundamental fields, which generalizes the one in \eqref{eq:superpotg2E0}. 

This theory also has a single $\mathbb{Z}_3^{[1]}$ 1-form symmetry coming from the unscreened diagonal combination of the center symmetries of each gauge node. In order to detect a possible anomaly for this symmetry, we consider again the version of the theory with gauge group $SU(3)^{2(g-1)}/\mathbb{Z}_3$ obtained by gauging it and we focus on the monopole operator with gauge fluxes $(-\frac{1}{3},-\frac{1}{3})$ for all of the $SU(3)$ gauge nodes. The contribution to the index of the bare monopole is
\be\label{Ibareg}
\zeta^2x^{-2(g-1)}\prod_{i=1}^{g-1}\left(\frac{y_{1}^{(i)}y_{2}^{(i)}}{z_{1}^{(i)}z_{2}^{(i)}}\right)^{2}\,,
\ee 
where $z_{1,2}^{(i)}$ and $y_{1,2}^{(i)}$ (with $i=1,\ldots, g-1$) are the fugacities of the $SU(3)$ gauge groups with positive and negative CS levels, respectively. Their presence indicates that the bare monopole we are considering is not gauge invariant. In order to obtain the basic gauge invariant monopole, we should dress this bare monopole with matter fields such that the resulting operator has the minimal R-charge possible in this flux sector. There are several different such possible dressings and each one corresponds to a different contribution to the same order of the index. We consider the dressed monopole obtained by dressing with $2(g-1)$ bifundamental anti-fermions with fugacities $\frac{z_{1}^{(i)}z_{2}^{(i)}}{y_{1}^{(i)}y_{2}^{(i)}}$, where $i$ takes values in all its domain $1,\ldots, g-1$. Their total contribution to the index is
\be
\prod_{i=1}^{g-1}\left(\frac{z_{1}^{(i)}z_{2}^{(i)}}{y_{1}^{(i)}y_{2}^{(i)}}x^{\frac{4}{3}}\right)^{2}
\ee 
and when combined with \eqref{Ibareg} we find that the basic gauge invariant monopole contributes at order $x^{\frac{2(g-1)}{3}}$. One can check that in the index of the $SU(3)^{2(g-1)}$ theory all the R-charges come in even integers and so we deduce that for generic $g$ the anomaly \eqref{eq:anomE0g2} becomes
\be\label{eq:anomE0g}
\frac{2\pi i(g-1)}{3}\int_{M_4} B_2 (C_1 (R)\text{ mod } 3)\,.
\ee 

Here we have not performed a complete analysis on all possible additional monopoles, and so cannot completely rule out the presence of a more general anomaly. However, this result matches with the explicit calculation for $g=2$, and as we shall next show it has a natural $5d$ interpretation.

\subsubsection*{Comparison with 5d}

We would like to relate the result we just found for the anomaly in the 3d model with some anomaly in the 5d $E_0$ SCFT via compactification. Recall that the latter has a $\mathbb{Z}_3^{[1]}$ 1-form symmetry, first found in \cite{Morrison:2020ool,Albertini:2020mdx}, and the superconformal symmetry. Here we shall only concentrate on internal symmetries, which eliminates most of the superconformal group save for the $SU(2)_R$ R-symmetry. Next consider the most general anomaly term one can write involving these symmetries:\footnote{Several remarks are in order. First the R-symmetry in general is $SU(2)$ not $SO(3)$, though in certain cases, like pure gauge theory, the central element may end up acting in the same way as fermion parity. We shall ignore this subtlety in what follows. Second, there is also the possibility of a Witten anomaly \cite{Witten:1982fp} in the $SU(2)_R$ R-symmetry \cite{Intriligator:1997pq}, and in fact this anomaly should exist for this theory. We will discuss the Witten anomaly for $SU(2)_R$ in detail in Section \ref{sec:wittenanom}.}
\be
\int_{M_6} \frac{\pi i \alpha}{9} (B_2)^3 + \frac{2\pi i \beta}{3} B_2( C_2 (R) \text{ mod } 3)\,,
\ee
where recall that $B_2$ is the background for the $\mathbb{Z}_3^{[1]}$ 1-form symmetry and $C_2 (R)$ is the second Chern class of the R-symmetry bundle in the fundamental representation, which is involved mod 3 in the anomaly. We point out that in the literature only the self-anomaly for the 1-form symmetry has been computed giving $\alpha=2$ \cite{Apruzzi:2021nmk}, while the value of $\beta$ is unknown.

We can next consider reducing the 5d SCFT on a genus $g$ Riemann surface by taking $M_6 = M_4 \times \Sigma_g$. To preserve supersymmetry we must couple the $U(1)_R$ Cartan of the $SU(2)_R$ R-symmetry to a background connection such that it cancels the curvature of the Riemann surface for some of the supercharges. At the level of characteristic classes, this sets
\be
C_2 (R) = - C^2_1 (R) + 2(1-g) t C_1 (R) + \cdots\,,
\ee
where $t$ is a unit 2-form on $\Sigma_g$ and the $\cdots$ indicate terms that are quadratic in $t$ and which are not relevant for our discussion. Additionally, we also have the option of turning on an holonomy in the 1-form symmetry on the surface and should also take into account that considering the background field $B_2$ for the 1-form symmetry to have support on one of the cycles of $\Sigma_g$ leads to degree one background fields for 0-form symmetries. We can take all this into account by writing
\be
\int_{\Sigma_g} B_2 = b \,,\qquad \int_{S^1_i \subset \Sigma_g} B_2 = A_1^{(i)}\,,
\ee       
where here $b$ is the value of the holonomy and $A_1^{(i)}\in H^1(M_3,\mathbb{Z}_3)$ are the background connections for the 0-form symmetries. Performing the integral we get
\be \label{3dAnomtermE0onSg}
\frac{2\pi i \alpha}{3} \sum_i \int_{M_4} A_1^{(2i)} A_1^{(2i+1)} B_2 + \int_{M_4} \frac{\pi i \alpha b}{3} (B_2)^2 + \int_{M_4}\frac{4\pi i \beta (1-g)}{3} B_2  (C_1 (R) \text{ mod }3)\,. 
\ee
Here we have ordered the $A_1^{(i)}$ fields such that $A_1^{(2i)}$ and $A_1^{(2i+1)}$ have non-trivial intersection. The first term describes a mixed anomaly between the 3d 1-form symmetry and the 0-form symmetries coming from the 5d 1-form symmetry wrapping intersecting 1-cycles. The second describes a self-anomaly of the 1-form symmetry. Finally the third term gives a mixed anomaly between the 1-form symmetry and the $U(1)_R$ R-symmetry. 

We want to compare this with what we observed in the 3d theory. In particular, we do not observe the 0-form symmetries coming from the 1-form symmetry on cycles and so we can't compare the first term of \eqref{3dAnomtermE0onSg}. The second term should vanish as there does not seem to be any obstruction to gauging the 1-form symmetry. This leads us to conclude that $b=0$. Finally, the third term precisely corresponds to the anomaly we observed in \eqref{eq:anomE0g} with $\beta = 1$.

This discussion suggests a mixed anomaly between the $\mathbb{Z}_3^{[1]}$ 1-form symmetry and the $SU(2)_R$ R-symmetry of the 5d $E_0$ SCFT
\be
\frac{2\pi i}{3}\int_{M_6} B_2( C_2 (R) \text{ mod } 3)\,,
\ee
It would be interesting to validate this result against some other computation of this anomaly directly in 5d.

\subsection{Anomalies between 0-form symmetries}
\label{sec:anom00form}

In this subsection we turn to the study of anomalies of the 3d models involving discrete aspects of their continuous 0-form symmetries. In the same spirit as before, we will do this by turning on non-trivial backgrounds for these symmetries and examining the properties of the new monopoles that are added by this procedure \cite{Tachikawa:2019dvq,Bergman:2020ifi,Beratto:2021xmn,Genolini:2022mpi,Bhardwaj:2022dyt,Mekareeya:2022spm,Bhardwaj:2023zix}. The main 3d models that we study are the ones arising from the compactification of the $E_{N_f+1}$ SCFTs on a genus 2 surface with no flux. In all of the cases in which we find a non-trivial anomaly in 3d, we show that this descends from a 5d anomaly which was not known before and so this provides new predictions for the 5d SCFTs.

\subsubsection{$E_{N_f+1}$ SCFT on a genus 2 surface for $0<N_f<7$}
\label{subsec:ENf_genus2}

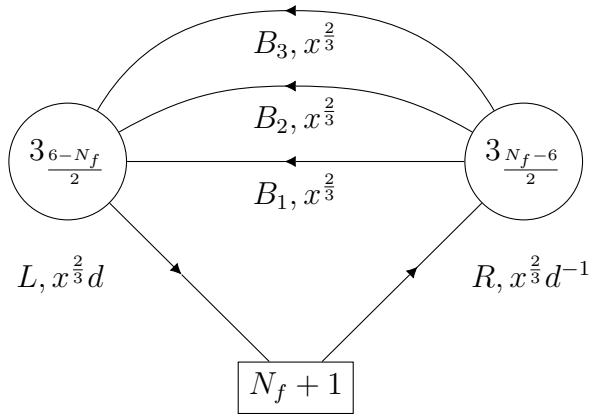
\begin{figure}[t]
\center
\begin{tikzpicture}[baseline=0, font=\scriptsize]
\node[draw, circle] (l) at (0,0) {\normalsize $3_{\frac{6-N_f}{2}}$};
\node[draw, circle] (r) at (6,0) {\normalsize $3_{\frac{N_f-6}{2}}$};
\node[draw, rectangle] (b) at (3,-3) {\normalsize $N_f+1$};

\draw[draw, solid,-<] (l)--(3,0);
\draw[draw, solid,-] (3,0)--(r);
\draw[draw, solid,-<] (l) edge [out=30,in=180,loop,looseness=1] (3,1);
\draw[draw, solid,-] (3,1) edge [out=0,in=150,loop,looseness=1] (r);
\draw[draw, solid,-<] (l) edge [out=60,in=180,loop,looseness=1] (3,2);
\draw[draw, solid,-] (3,2) edge [out=0,in=120,loop,looseness=1] (r);
\draw[draw, solid,->] (l)--(1.5,-1.5);
\draw[draw, solid,-] (1.5,-1.5)--(b);
\draw[draw, solid,-<] (r)--(4.5,-1.5);
\draw[draw, solid,-] (4.5,-1.5)--(b);

\node[] at (3,-0.4) {\normalsize $B_1,x^{\frac{2}{3}}$};
\node[] at (3,1-0.4) {\normalsize $B_2,x^{\frac{2}{3}}$};
\node[] at (3,2-0.4) {\normalsize $B_3,x^{\frac{2}{3}}$};
\node[] at (-0.1,-1.5) {\normalsize $L,x^{\frac{2}{3}}d$};
\node[] at (6.1,-1.5) {\normalsize $R,x^{\frac{2}{3}}d^{-1}$};
\end{tikzpicture}
\caption{The $3d$ model for the compactification of the 5d rank 1 $E_{N_f+1}$ SCFT on a genus 2 surface with no flux.}
\label{fig:Genus2dual}
\end{figure}

We want to study anomalies involving discrete aspects of the 0-form symmetries of the 3d theories obtained by compactifying the 5d $E_{N_f+1}$ SCFTs on a genus 2 surface with no flux. These models, which were discussed in detail in \cite{Sacchi:2023rtp}, are summarized in Figure \ref{fig:Genus2dual}. The superpotential is of the
form
\be\label{superpotgenus2}
\mathcal{W}=\sum_{i,j=1}^3B_i(B_j)^2+\sum_{i=1}^3LB_iR\,.
\ee

This theory has an $\mathfrak{su}(N_f+1)\oplus \mathfrak{u}(1)_d$ manifest global symmetry which gets enhanced to $\mathfrak{e}_{N_f+1}$ in the IR. The precise global structure will be specified for each value of $N_f$ separately and as we will see in some cases it will allow us to turn on fractional background magnetic fluxes for this symmetry and study the contribution to the index of the associated monopole operators. In order to make these fractional fluxes consistent, they should be compensated by fractional fluxes also for the $SU(3)\times SU(3)$ gauge symmetry. Observe that this is not always possible for the $\mathfrak{su}(N_f+1)$ symmetry since its center is $\mathbb{Z}_{N_f+1}$ while the center of $SU(3)$ is $\mathbb{Z}_3$ and $\textrm{GCD}(N_f+1,3)\neq 1$ only for $N_f=2,5$. This suggests that there cannot be any anomaly involving this symmetry except for these two values of $N_f$. On the other hand, there could be anomalies for $\mathfrak{u}(1)_d$, which we can detect by turning on a flux for it. The anomalies detected from these background monopoles might then imply an anomaly for the enhanced $\mathfrak{e}_{N_f+1}$ symmetry, which should descend from some 5d anomaly. 

\subsubsection*{\boldmath$N_f=6$}

In this case the actual global symmetry is really $SU(7)\times U(1)_d$. This is because we cannot use a transformation by the center of the $SU(3)$ gauge groups to reabsorbe a transformation by the center of the $SU(7)$ flavor group. Hence, we cannot turn on any background monopole with fractional flux and there is no discrete anomaly involving the $SU(7)$ symmetry. On the other hand, fractional fluxes for $U(1)_d$ are allowed. Indeed, one can check from the index computation of \cite{Sacchi:2023rtp} that the minimal charge for this symmetry appearing in the spectrum of the theory is 3, and so we can turn on a flux $\frac{1}{3}$. This has to be compensated with a fractional flux also for the $SU(3)\times SU(3)$ gauge group of the form $\left(\frac{1}{3},\frac{1}{3};\frac{1}{3},\frac{1}{3}\right)$. The contribution to the index from such a monopole sector turns out to be zero, indicating that there is no anomaly involving $U(1)_d$.

\subsubsection*{\boldmath$N_f=5$}

For $N_f=5$ the actual flavor symmetry is
\be
\frac{SU(6)}{\mathbb{Z}_3}\times U(1)_d\,.
\ee
That the $\mathbb{Z}_3$ subgroup of the $SU(6)$ center acts trivially on the spectrum can be seen from the index computation of \cite{Sacchi:2023rtp}, but again it can also be easily understood from the fact that such a $\mathbb{Z}_3$ transformation can be re-absorbed by a transformation for the center of the gauge group. We can then consider a flux for the gauge$\times$flavor symmetry of the form
\be
\left(\frac{1}{3},\frac{1}{3};\frac{1}{3},\frac{1}{3};\frac{1}{3},\frac{1}{3},\frac{1}{3},-\frac{1}{3},-\frac{1}{3},-\frac{1}{3}\right)\,,
\ee
where the first four entries are the $SU(3)\times SU(3)$ gauge flux while the last six entries are the $SU(6)$ flavor flux in an overcomplete parametrization where their sum is zero.
This monopole sector gives a non-trivial contribution to the index
\be
3x^{\frac{4}{3}}\,.
\ee
The coefficient of $x$ not being an integer multiple of 2 as in the result for the index with no background flux (see eq.~(4.15) of \cite{Sacchi:2023rtp}) suggests the presence of a mixed anomaly between the $SU(6)/\mathbb{Z}_3$ flavor symmetry and the $U(1)_R$ R-symmetry
\be
\frac{4\pi i}{3}\int_{M_4} w_2 (SU(6)/\mathbb{Z}_3) \left(C_1 (R)\text{ mod }3\right)	\,.
\ee
The flux we considered is also a fractional flux for the enhanced $E_6/\mathbb{Z}_3$ symmetry and so our result suggests that also this enhanced symmetry should have an anomaly with the R-symmetry
\be\label{eq:anomE6g23d}
\frac{4\pi i}{3}\int_{M_4} w_2 (E_6/\mathbb{Z}_3) \left(C_1 (R)\text{ mod }3\right)	 \,.
\ee
Here we used the embedding of $\mathfrak{su}(6)\times \mathfrak{su}(2)$ inside $\mathfrak{e}_6$
\be\label{eq:embSU2SU6inE6}
{\bf 27}_{\mathfrak{e}_6} \rightarrow ({\bf 2}_{\mathfrak{su}(2)},{\bf 6}_{\mathfrak{su}(6)}) + ({\bf 1}_{\mathfrak{su}(2)},\overline{\bf 15}_{\mathfrak{su}(6)})\,,
\ee
which suggests that the non-trivial $E_6/\mathbb{Z}_3$ bundles should reduce to non-trivial $SU(6)/\mathbb{Z}_3$ bundles.

We can also study fractional fluxes for $U(1)_d$. As in the previous example, we can turn on a gauge$\times$flavor flux of the form $\left(\frac{1}{3},\frac{1}{3},\frac{1}{3},\frac{1}{3};\frac{1}{3}\right)$, which gives the following contribution to the index:
\be
3x^{\frac{4}{3}}\,.
\ee
This would suggest an anomaly between $U(1)_d$ and the $U(1)_R$ R-symmetry. Nevertheless, it is not a true anomaly since it can be removed by adding a suitable local counter-term. Such a counter-term is a mixed CS coupling between the two abelian symmetries at level 2\footnote{One might be worried that such a counter-term would break parity and time reversal. However, note that the theory already contains CS terms so these symmetries are actually not present. Nevertheless, we can combine them with quiver reflection to get a symmetry, but under quiver reflection we have that the background $U(1)_{d}$ connection transforms as $A_{d}\rightarrow -A_{d}$ so the counter-term preserves these generalized parity  and time reversal symmetries.}
\be
S_{\text{c.t.}}=\frac{2}{4\pi}\int_{M_3}A_d\mathrm{d}A_R+A_R\mathrm{d}A_d\,,
\ee
where $A_d$ and $A_R$ are the background fields for $U(1)_d$ and $U(1)_R$, which contributes to the index with a prefactor $x^{2n_d}$ where $n_d$ is the $U(1)_d$ flux. Repeating the previous computation in the presence of such a counter-term we indeed find
\be
3x^{2}+(1+{\bf 35}_{SU(6)})x^4+\cdots\,,
\ee
which doesn't indicate any anomaly. This is compatible with the anomaly \eqref{eq:anomE6g23d} we wrote above, since we see from the embedding \eqref{eq:embSU2SU6inE6} that a non-trivial $E_6/\mathbb{Z}_3$ bundle reduces to a trivial bundle for the $SU(2)$ of which $U(1)_d$ is the Cartan.

The anomaly \eqref{eq:anomE6g23d} can be obtained from the 5d anomaly
\be
\frac{4\pi i}{3}\int_{M_6} w_2 (E_6/\mathbb{Z}_3) \left(C_2 (R)\text{ mod }3\right) \,,
\ee
where the derivation is completely analogous to the one we did in Subsection \ref{subsec:E0g} but with  the background field $B_2$ for the 1-form symmetry being replaced by $w_2 (E_6/\mathbb{Z}_3)$. Here we further assume that $w_2 (E_6/\mathbb{Z}_3)$ has no support on the compact surface, which should be true for the case at hand given that there is no flux for the $E_6/\mathbb{Z}_3$ symmetry through the Riemann surface. 

\subsubsection*{\boldmath$N_f=4$}

For $N_f=4$, similarly to the $N_f=6$ case, we cannot turn on consistently any fractional flux for the non-abelian symmetry since this is $SU(5)$. This means that there can't be any anomaly for this symmetry. On the other hand, we can in principle consider a fractional flux for $U(1)_d$. Again, from the index computation of \cite{Sacchi:2023rtp} one can see that the minimal charge under this symmetry is 3 and so we can consider the gauge$\times$flavor flux $\left(\frac{1}{3},\frac{1}{3},\frac{1}{3},\frac{1}{3};\frac{1}{3}\right)$, where the first four entries are for the $SU(3)\times SU(3)$ gauge symmetry while the last one is for $U(1)_d$. This gives the non-trivial contribution to the index
\be
x^{-\frac{1}{3}}\,,
\ee
but again this doesn't indicate any anomaly since as before it can actually be removed by a local counter-term. In this case this is taken to be a mixed CS coupling between $U(1)_d$ and the $U(1)_R$ R-symmetry at level 7 that contributes to the index as a prefactor of $x^{7n_d}$, so that now the index computation gives
\be
x^2+(3+{\bf 24}_{SU(5)})x^4+\cdots\,.
\ee
Overall, we do not find any anomaly in this case.

\subsubsection*{\boldmath$N_f=3$}

Also for $N_f=3$ no fractional flux for the non-abelian symmetry is allowed since this is $SU(4)$, but we can still consider the flux $\left(\frac{1}{3},\frac{1}{3},\frac{1}{3},\frac{1}{3};\frac{1}{3}\right)$ for the gauge$\times$flavor $U(1)_d$ symmetry. This gives a non-trivial contribution to the index
\be
3+\left(1+{\bf 15}+{\bf 20}\right)x^2\,,
\ee
which nevertheless, after comparing with the index calculation of \cite{Sacchi:2023rtp} for zero background flux for $U(1)_d$, suggests no anomaly for this symmetry. Like the previous case, we do not find any anomaly.

\subsubsection*{\boldmath$N_f=2$}

For $N_f=2$ the actual global symmetry is
\be
\frac{SU(3)}{\mathbb{Z}_3}\times U(1)_d\,,
\ee
as one can see either from the index computation of \cite{Sacchi:2023rtp} or from the fact that such a $\mathbb{Z}_3$ transformation can be re-absorbed by a transformation for the center of the gauge group. Moreover, unlike in the previous cases, they don't combine to form a single larger symmetry, since the full enhanced symmetry is $\mathfrak{su}(3)\times \mathfrak{su}(2)_d$ where $\mathfrak{su}(2)_d$ is enhanced solely from $\mathfrak{u}(1)_d$. 

We start by considering possible anomalies for the non-abelian part, so we consider the gauge$\times$flavor flux $\left(\frac{1}{3},\frac{1}{3},\frac{1}{3},\frac{1}{3};\frac{1}{3},\frac{1}{3}\right)$, where as usual the first four entries correspond to the $SU(3)\times SU(3)$ gauge flux while the last two to the $SU(3)$ flavor flux. This gives the following contribution to the index:
\be
10x^{\frac{4}{3}}\,.
\ee
This result should be compared with the index calculation of \cite{Sacchi:2023rtp} for zero background $SU(3)$ flux, where all the R-charges are even integers. Hence, this computation indicates an anomaly between the $SU(3)/\mathbb{Z}_3$ and the $U(1)_R$ symmetries
\be\label{eq:anomE3g23d}
\frac{4\pi i}{3}\int_{M_4} w_2 (SU(3)/\mathbb{Z}_3) \left(C_1 (R)\text{ mod }3\right)\,.
\ee

For the abelian part, we consider the usual monopole $\left(\frac{1}{3},\frac{1}{3},\frac{1}{3},\frac{1}{3};\frac{1}{3}\right)$, whose corresponding contribution to the index is
\be
6x^{\frac{1}{3}}\,.
\ee
Again, this doesn't indicate any anomaly since after adding a local counter-term which is a mixed CS coupling between $U(1)_d$ and $U(1)_R$ at level 5 that contributes to the index with a prefactor of $x^{5n_d}$, the same index computation gives
\be
6x^2+({\bf 8}_{SU(3)}-5)x^4+\cdots\,.
\ee

Overall, we only find the anomaly \eqref{eq:anomE3g23d} between $SU(3)/\mathbb{Z}_3$ and $U(1)_R$, which using similar arguments as before can be shown to descend from the 5d anomaly
\be
\frac{4\pi i}{3}\int_{M_6} w_2 (SU(3)/\mathbb{Z}_3) \left(C_2 (R) \text{ mod }3\right)\,.
\ee

\subsubsection*{\boldmath$N_f=1$}

In this case we cannot consider any fractional flux for the non-abelian $SU(2)$ symmetry. For the $U(1)_d$ symmetry it is not clear which fluxes are allowed, since the computation of the index of \cite{Sacchi:2023rtp} without background flux shows no appearance of the fugacity $d$ up to order $x^3$. Nevertheless, we do not expect any anomaly for $U(1)_d$ since as in the previous examples it can always be removed by adding a local counter-term.

\section{5d Witten anomaly and 3d R-symmetry parity anomaly}
\label{sec:wittenanom}

\subsection{General discussion}

In this section we study the Witten anomaly \cite{Witten:1982fp} for the $SU(2)_R$ R-symmetry of 5d theories and its 3d reduction. As $\pi_5 (USp(2N)) = \mathbb{Z}_2$, we can have a Witten anomaly for $USp(2N)$ groups, and in particular for $SU(2)$ groups. In the case at hand, we always have an $SU(2)$ global symmetry, which is the R-symmetry of the 5d SCFT, and we can consider whether it suffers from a Witten anomaly.\footnote{In certain cases we also have other $SU(2)$ flavor symmetries that we can consider. However, string theory constructions suggest these are all gaugeable and so, non-anomalous.} 

For 5d gauge theories, the possible presence of a Witten anomaly for $SU(2)_R$ can be read off from the matter content. Specifically, a 5d $SU(2)$ global symmetry has a Witten anomaly whenever it sees an odd number of symplectic Majorana fermions \cite{Intriligator:1997pq}. Both the half-hypermultiplets and the vector multiplets contain such a fermion, with the one in the vector being charged under the $SU(2)_R$ R-symmetry, while the one in the half-hyper being neutral. As such we see that only the vector multiplets contribute to such an anomaly, which is present whenever the number of vectors is odd.

Next consider generic 5d SCFTs. We first note that the Witten anomaly should match across RG flows that do not break the $SU(2)_R$ R-symmetry. This includes both 5d real mass deformations and going on the Coulomb branch. The latter is of particular interest, as it just converts the 5d SCFT to an abelian gauge theory of $r$ free vectors where $r$ is the rank of the theory, that is the real dimension of its Coulomb branch. The R-symmetry of the $U(1)^r$ theory on a generic point of the Coulomb branch has a Witten anomaly whenever $r$ is odd, so we conclude that for any 5d SCFT with a Coulomb branch, the $SU(2)_R$ R-symmetry has a Witten anomaly whenever the rank of the theory is odd and is free of such anomaly if the rank is even.

We can next wonder what happens to the Witten anomaly anomaly under dimensional reduction to 3d. Here we shall specifically have in mind the reduction with twisting and fluxes in the global symmetry such that we in general get a 3d $\mathcal{N}=2$ theory. The (trial) R-symmetry of the latter is $U(1)_R$ which is the Cartan of the 5d $SU(2)_R$ R-symmetry. As such we can approach this question in steps, by first breaking the 5d $SU(2)_R$ to a 5d $U(1)_R$ symmetry and then consider the reduction to a 3d $U(1)_R$ symmetry.  

Consider $n$ doublets of symplectic-Majorana fermions in the fundamental of an $SU(2)$ global symmetry. From the preceding discussion, this symmetry carries a Witten anomaly whenever $n$ is odd. Now consider its $U(1)$ Cartan. The $n$ doublets now become $n$ charge $1$ Dirac fermions under the $U(1)$, and it is known that the $U(1)$ has a parity anomaly whenever $n$ is odd. As such, we observe that the Witten anomaly for the $SU(2)$ implies a parity anomaly for its $U(1)$ Cartan. This is quite reasonable as both originate from a lack of gauge invariance under large gauge transformations.\footnote{The previous argument assumes a Lagrangian description. We expect that it should be possible to show this without reference to a specific Lagrangian though we shall not pursue it here.}

The parity anomaly can be described by the anomaly theory 
\be
\frac{\pi i}{6}\int_{M_6} (C_1 (R)\text{ mod 2})^3\,,
\ee
where $C_1 (R)$ is the first Chern class of the $U(1)_R$ Cartan of the 5d $SU(2)_R$ R-symmetry. We can next consider integrating this anomaly term to find the resulting 3d anomaly.\footnote{One can consider doing the same also for the Witten anomaly for an unbroken $SU(2)$. However, here the anomaly theory is the mod 2 reduction of the 6d Dirac operator, which is harder to work with, and so we shall not consider this here. We are grateful to Zohar Komargodski and Yichul Choi for pointing this out to us.} Here we only care about the flux in $U(1)_R$ which is dictated by the desire to preserve SUSY to be
\be
\int_{\Sigma_g} C_1^{2d} (R) = g-1\,,
\ee   
with $g$ the genus of the surface $\Sigma_g$ and where we took $C_1 (R) = C^{2d}_1 (R) + C^{3d}_1 (R)$, with $C^{2d}_1$ standing for the part of the Chern class on the 2d surface and $C^{3d}_1$ the part on the 3d spacetime. We then have
\be
\frac{\pi i}{6}\int_{M_6} (C_1 (R)\text{ mod 2})^3 \rightarrow \frac{(g-1)\pi i}{2}\int_{M_4} (C_1^{3d} (R)\text{ mod 2})^2 \,,
\ee
where as usual $M_6=M_4\times\Sigma_g$. Similarly to 5d, the 3d parity anomaly for a $U(1)$ symmetry can be associated with the anomaly theory 
\be
\frac{\pi i}{2}\int (C_1 (U(1))\text{ mod }2)^2\,.
\ee
We therefore observe that the 5d Witten anomaly for $SU(2)_R$ reduces to a parity anomaly in the $U(1)_R$ Cartan in 3d whenever the genus is even.\footnote{Once we compactify, the 5d parity transformation breaks into parity along the 2d compact space and parity along the 3d non-compact space. Note that the former will in general be broken by the flux, but it is only the latter one, which becomes the parity transformation of the 3d theory, that would be of interest to us here.}

Our main goal for this section is to test this in a variety of examples. We by now have several 3d models believed to describe the compactification of 5d SCFTs on various Riemann surfaces \cite{Sacchi:2021afk,Sacchi:2021wvg,Sacchi:2023rtp}, some of which we encountered in the previous section, that can be used to perform such tests by computing the parity anomaly for $U(1)_R$ in 3d and trying to match it with the Witten anomaly for $SU(2)_R$ in 5d. The latter can be easily deduced by just looking at the rank of the 5d SCFT as previously reviewed, while the former can be obtained by using the fact that a $U(1)$ flavor symmetry in 3d has a parity anomaly if $\sum_i q^2_i = \text{odd}$, where the sum goes over all the Dirac fermions charged with charge $q_i$ under said $U(1)$ and we assumed a normalization of the $U(1)$ charge is chosen such that the minimal charge is $1$. We shall next consider various cases.

\subsection{Compactifications on tori}    

There are several known examples of torus compactifications, where since $g-1=0$ by our previous argument we expect no parity anomaly for $U(1)_R$. These include the compactifications of the rank $1$ $E_{N_f+1}$ theories, of which we discussed the case $N_f=0$ in Subsection \ref{subsec:E1torus}, as well as their higher rank generalizations to compactifications of the 5d SCFTs UV completing 5d $SU$ type gauge theories with fundamental matter \cite{Sacchi:2021afk,Sacchi:2021wvg}, which we shall consider in turn. 

The torus theories can be constructed by gluing a number $F$ of elementary building blocks corresponding to tube theories. We can then compute the contribution to the $U(1)_R$ parity anomaly coming from each component tube and from the gluing to find the anomaly of the torus model.

Let us briefly review how this gluing works (we refer to \cite{Sacchi:2021afk,Sacchi:2021wvg,Sacchi:2023rtp} for more details). The tube theory is equipped with two global symmetries that correspond to each of the punctures, whose group corresponds with the gauge group of the gauge theory phase of the 5d SCFT. Moreover, for each puncture there is a set of operators, sometimes referred to as ``moment maps", which transform under the puncture symmetry and under a subgroup of the 5d global symmetry, which depends on the value of the flux and on the type of puncture. 

Let us now describe the gluing of two tubes along one puncture from each. We shall denote by $G_L$, $G_R$ the symmetries of the two punctures of one tube and by $G_L'$, $G_R'$ those of the other. Similarly, we denote by $M_L$, $M_R$ the moment map operators of one tube and those of the other by $M_L'$, $M_R'$. Then, if the puncture $R$ of one tube and the puncture $L'$ of the other are of the same type, so in particular $G_R=G_L'$, we can glue them as follows:
\begin{enumerate}
\item Identify the two symmetries $G=G_R=G_L'$ and possibly introduce chiral fields charged under $G$ with superpotential terms coupling them with the fields $M_R$ and $M_L'$, as will be specified momentarily;
\item Gauge $G$.
\end{enumerate}

Step 1.~gives some freedom in the fields we introduce. In particular, we have two opposite possibilities that give rise to two different types of gluing:
\begin{itemize}
\item \emph{$S$-gluing}: we do not introduce any extra chiral and we turn on the superpotential interaction
\be\label{eq:WS}
\gd\mathcal{W}_S=M_R\cdot M_L'\,.
\ee
\item \emph{$\Phi$-gluing}: introduce extra chirals $\Phi$ and the interaction
\be
\gd\mathcal{W}_\Phi=\Phi\cdot\left(M_R-M_L'\right)\,.
\ee
\end{itemize}
There can also be intermediate situations where we introduce only a subset of the chirals $\Phi$ which interact only with some of the components of $M_R$ and $M_L'$, while for the others we have the interaction \eqref{eq:WS} of the $S$-gluing.

Depending on the type of gluing, the fluxes of the glued tubes compose in different ways. This will not be relevant for our discussion of the Witten anomaly and so we will neglect this aspect, referring the reader to \cite{Sacchi:2021afk,Sacchi:2021wvg,Sacchi:2023rtp} for the details. What is instead crucial for us is that the moment map operators have R-charge 1 under the $U(1)_R$ Cartan of the 5d $SU(2)_R$ R-symmetry. Hence, the R-charge of the chirals introduced when performing the $\Phi$-gluing is also 1. This in particular means that there is no contribution to the $U(1)_R$ parity anomaly from these fields, since their fermionic components have 0 R-charge. Nevertheless, there will still be contributions to the anomaly from the gauginos that we introduce in the gauging, but this is the same for $S$ and $\Phi$-gluing. Hence, there is no dependence of the result for the $U(1)_R$ parity anomaly on the type of gluing performed. 

\subsubsection*{Rank $1$ $E_{N_f+1}$ theories} 

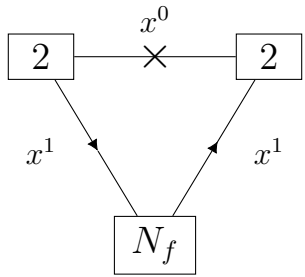
\begin{figure}[t]
\center
\begin{tikzpicture}[baseline=0, font=\scriptsize]
\node[draw, rectangle] (l) at (0,0) {\large $\,\, 2\,\,$};
\node[draw, rectangle] (r) at (3,0) {\large $\,\, 2\,\,$};
\node[draw, rectangle] (b) at (1.5,-2.5) {\large $\, N_f\,$};
\draw[draw, solid,-] (l)--(r);

\draw[draw, solid,->] (l)--(0.75,-1.25);
\draw[draw, solid,-] (0.75,-1.25)--(b);
\draw[draw, solid,-<] (r)--(2.25,-1.25);
\draw[draw, solid,-] (2.25,-1.25)--(b);

\node[thick,rotate=0] at (1.5,0) {\Large$\times$};

\node[] at (1.5,0.5) {\normalsize $x^0$};
\node[] at (0,-1.25) {\normalsize $x^1$};
\node[] at (3,-1.25) {\normalsize $x^1$};
\end{tikzpicture}
\caption{The 3d $\mathcal{N}=2$ Lagrangian of \cite{Sacchi:2021afk} for the compactification of the 5d rank 1 $E_{N_f+1}$ SCFT on a tube with flux. The cross denotes a singlet that couples to the meson constructed from the bifundamental.}
\label{fig:3dtube}
\end{figure} 

We start considering the torus compactification of the rank 1 $E_{N_f+1}$ SCFTs with flux. We refer the reader to \cite{Sacchi:2021afk} for more details on these models, while here we will just summarize the aspects needed for the computation of the anomaly.

The basic tube is given as a collection of chiral fields, that can be arranged as bifundamentals of the $SU(2)\times SU(2)\times SU(N_f)$ global symmetry and a singlet \cite{Sacchi:2021afk}, see Figure \ref{fig:3dtube}. Their R-charges under the $U(1)_R$ Cartan of the 5d $SU(2)_R$ R-symmetry are $1$ for the $SU(2)\times SU(N_f)$ bifundamentals, $0$ for the $SU(2)\times SU(2)$ bifundamentals and $2$ for the singlet, as summarized in the figure. In this case the symmetry carried by each opuncture is $SU(2)$ and the moment map operators $M_L$ and $M_R$ are the $SU(2)\times SU(N_f)$ bifundamentals, since the gauge theory associated with the 5d SCFT is $SU(2)$ with $N_f$ flavors. For all $N_f$ we have that the contribution from a single tube to the $U(1)_R$ parity anomaly is
\be
\mathrm{Tr}(U(1)^2_R)_{\text{Tube}} = 4 \times (-1)^2 + 1^2 = 5\,. 
\ee

As reviewed above, the torus can be constructed by gluing $F$ copies of this tube, which is done in field theory by gauging the $SU(2)$ puncture symmetries with $N_f$ fundamental chiral fields of R-charge 1.\footnote{In this set-up the $S$-gluing is not allowed. This can be understood from the fact that the resulting flux would be zero and the compactification on a torus with no flux is not contemplated by our construction.}
With this in mind, we can easily compute the anomaly of the torus by adding $F$ times that of the tube and the contribution of the $F$ $SU(2)$ gauginos
\be
\mathrm{Tr}(U(1)^2_R)_{\text{Torus}} = 5F+3F = 8F\,. 
\ee
This number is even for any integer $F$ so we indeed have no anomaly as expected.

\subsubsection*{Higher rank cases}

\begin{figure}[t]
\center
\begin{tikzpicture}[baseline=0, font=\scriptsize]
\node[draw, rectangle] (l) at (0,0) {\large $N+1$};
\node[draw, rectangle] (r) at (4,0) {\large $N+1$};
\node[draw, rectangle] (b) at (2,-2.5) {\large $\, p\,$};
\node[draw, rectangle] (t) at (2,2.5) {\large $N_f-p$};

\draw[draw, solid,->] (l)--(2,0);
\draw[draw, solid,-] (2,0)--(r);

\draw[draw, solid,->] (l)--(1,-1.25);
\draw[draw, solid,-] (1,-1.25)--(b);
\draw[draw, solid,-<] (r)--(3,-1.25);
\draw[draw, solid,-] (3,-1.25)--(b);

\draw[draw, solid,-<] (l)--(1,1.25);
\draw[draw, solid,-] (1,1.25)--(t);
\draw[draw, solid,->] (r)--(3,1.25);
\draw[draw, solid,-] (3,1.25)--(t);

\node[thick,rotate=0] at (1.4,0) {\Large$\times$};

\node[] at (2.3,0.5) {\normalsize $x^0$};
\node[] at (0,-1.25) {\normalsize $x^1$};
\node[] at (4,-1.25) {\normalsize $x^1$};
\node[] at (0,1.25) {\normalsize $x^1$};
\node[] at (4,1.25) {\normalsize $x^1$};
\end{tikzpicture}
\caption{The general structure of the 3d $\mathcal{N}=2$ tube theory of \cite{Sacchi:2021wvg} associated to 5d SCFTs that UV complete $SU(N+1)_k+N_fF$ gauge theories with two $SU(N+1)$ punctures.}
\label{fig:3dtubehigherrank}
\end{figure}
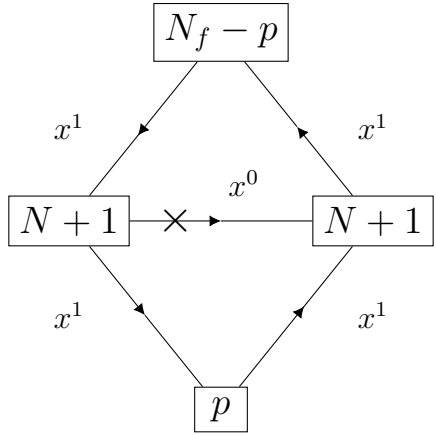 

We next consider the cases where the 5d SCFT is the one that UV completes a 5d $SU(N+1)$ gauge theory with $N_f$ fundamental hypermultiplets and Chern--Simons level $k$. The corresponding tubes were determined in \cite{Sacchi:2021wvg} for some range of the parameters $N$, $N_f$ and $k$. There are two types of tubes, one involving two $SU(N+1)$ punctures and one involving an $SU(N+1)$ and a $USp(2N)$ puncture. This is thanks to the fact that some of these 5d SCFTs admit multiple gauge theory descriptions. We shall next consider both possibilities in turn.

Let us begin with the case with two $SU(N+1)$ punctures, which we summarize in Figure \ref{fig:3dtubehigherrank}. Similarly to the rank $1$ case, the tube is made from bifundamentals of the $SU(N+1)\times SU(N+1)\times SU(N_f-p)\times SU(p)$ global symmetry, for some number $p$ depending on $k$, and a singlet. Their R-charges under the $U(1)_R$ Cartan of the 5d $SU(2)_R$ R-symmetry are $1$ for the $SU(N+1)\times SU(N_f-p)$ and $SU(N+1)\times SU(p)$ bifundamentals, $0$ for the $SU(N+1)\times SU(N+1)$ bifundamentals and $2$ for the singlet, as summarized in the figure. 

The gluing is done by gauging the $SU(N+1)$ global symmetry with $N_f$ fundamental chiral fields, all with R-charge $1$. As such we again observe that the anomaly coefficient is independent of $N_f$, and is given by
\be
\mathrm{Tr}(U(1)^2_R)_{\text{Torus}} = ((N+1)^2+1)F+((N+1)^2-1)F = 2(N+1)^2F \,, 
\ee
which is again even for all $N$.

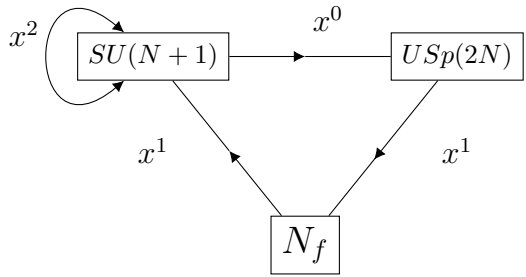
\begin{figure}[t]
\center
\begin{tikzpicture}[baseline=0, font=\scriptsize]
\node[draw, rectangle] (l) at (0,0) {\fontsize{10pt}{10pt}\selectfont $SU(N+1)$};
\node[draw, rectangle] (r) at (4,0) {\fontsize{10pt}{10pt}\selectfont $USp(2N)$};
\node[draw, rectangle] (b) at (2,-2.5) {\large $N_f$};

\draw[draw, solid,->] (l)--(2,0);
\draw[draw, solid,-] (2,0)--(r);

\draw[draw, solid,->] (b)--(1,-1.25);
\draw[draw, solid,-] (1,-1.25)--(l);
\draw[draw, solid,-<] (b)--(3,-1.25);
\draw[draw, solid,-] (3,-1.25)--(r);

\draw[black,solid,<->] (l) edge [out=220,in=140,loop,looseness=7]  (l);

\node[] at (2.3,0.5) {\normalsize $x^0$};
\node[] at (0,-1.25) {\normalsize $x^1$};
\node[] at (4,-1.25) {\normalsize $x^1$};
\node[] at (-1.7,0.3) {\normalsize $x^2$};
\end{tikzpicture}
\caption{The general structure of the 3d $\mathcal{N}=2$ tube theory of \cite{Sacchi:2021wvg} associated to 5d SCFTs that UV complete $SU(N+1)_k+N_fF$ gauge theories with one $SU(N+1)$ and one $USp(2N)$ puncture.}
\label{fig:3dtubehigherrank2}
\end{figure}

We next consider the tube with an $SU(N+1)$ and a $USp(2N)$ puncture. The tube, schematically represented in Figure \ref{fig:3dtubehigherrank2}, is made from an $SU(N+1)\times USp(2N)$ bifundamental of $U(1)_R$ R-charge $0$, an $SU(N+1)$ antisymmetric chiral of R-charge $2$ and several $SU(N+1)$ and $USp(2N)$ fundamental flavors of R-charge $1$ that as usual will not matter in this discussion since their R-charge is 1. 

The gluing is done via gauging the puncture symmetries, $SU(N+1)$ or $USp(2N)$ depending on the punctures, up to chiral fields of R-charge $1$ which again are not relevant. As such we see that
\bea
\mathrm{Tr}(U(1)^2_R)_{\text{Torus}} & = & \left(2N(N+1)+\frac{1}{2}N(N+1)\right)F+\left((N+1)^2-1 + N(2N+1)\right)\frac{F}{2} \nonumber \\ & = & 4N(N+1)F \,, 
\eea
which is again even. We note that here $F$ must be even for us to be able to glue all tubes into a torus.

Finally, we can consider the case where we take $2x$ $SU(N+1)\times USp(2N)$ tubes and $y$ $SU(N+1)\times SU(N+1)$ tubes. We then have
\bea
 \mathrm{Tr}(U(1)^2_R)_{\text{Torus}} & = & 2x(2N(N+1)+\frac{1}{2}N(N+1))+y((N+1)^2+1) \\ \nonumber & + & y N(2N+1) + 2x((N+1)^2-1) + N (7N+9)x + y (3N^2+3N+2) \,. 
\eea
We see that it is again even for every integer $x$, $y$ and $N$.

\subsection{Compactifications on higher genus surfaces}

We can also consider the reduction on higher genus surfaces for the rank $1$ Seiberg $E_{N_f+1}$ theories. Here we expect the anomaly coefficient to be proportional to $(g-1)$ and so to be present for even genera. We refer the reader to \cite{Sacchi:2023rtp} for more details on these models, while here we will just summarize the aspects needed for the computation of the anomaly.

\begin{figure}[t]
\center
\begin{tikzpicture}[baseline=0, font=\scriptsize]
\node[draw, circle] (c) at (0,0) {\large $ 3_{\frac{6-N_f}{2}}$};
\node[draw, rectangle] (t) at (0,3) {\large $N_f-2$};
\node[draw, rectangle] (l) at (-3,0) {\large $\,\, 2\,\,$};
\node[draw, rectangle] (r) at (3,0) {\large $\,\, 2\,\,$};
\node[draw, rectangle] (b) at (0,-3) {\large $\,\, 2\,\,$};
\draw[draw, solid,->] (c)--(0,1.5);
\draw[draw, solid,-] (0,1.5)--(t);
\draw[draw, solid,-<] (c)--(-1.5,0);
\draw[draw, solid,-] (-1.5,0)--(l);
\draw[draw, solid,-<] (c)--(1.5,0);
\draw[draw, solid,-] (1.5,0)--(r);
\draw[draw, solid,-<] (c)--(0,-1.5);
\draw[draw, solid,-] (0,-1.5)--(b);

\node[] at (1.9,1.5) {\normalsize $x^{\frac{2}{3}}$};
\node[] at (-1.7,-0.6) {\normalsize $x^{\frac{1}{3}}$};
\node[] at (1.7,-0.6) {\normalsize $x^{\frac{1}{3}}$};
\node[] at (-1.2,-1.8) {\normalsize $x^{\frac{1}{3}}$};

\end{tikzpicture}
\caption{The 3d $\mathcal{N}=2$ Lagrangian of \cite{Sacchi:2023rtp} for the compactification of the 5d rank 1 $\mathfrak{e}_{N_f+1}$ SCFT on a three punctured sphere with flux.}
\label{fig:3dtrinion}
\end{figure}
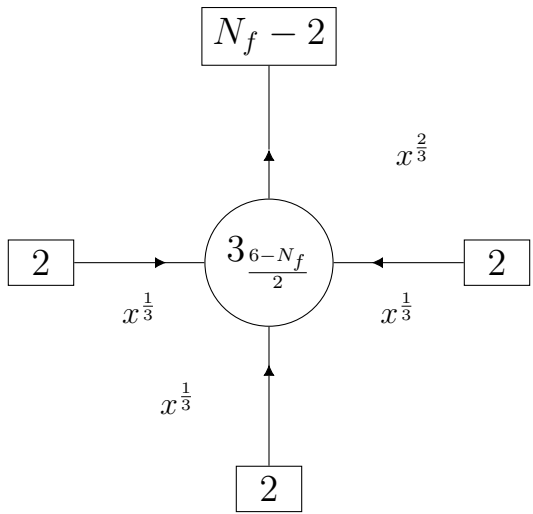

We can establish this as follows. First we consider the theories made from $S$-gluing various copies of the trinion theory proposed in \cite{Sacchi:2023rtp}, which we report in Figure \ref{fig:3dtrinion}. The gluing prescription works exactly as in the case of the torus described above, but the only difference is that to get higher genus surfaces the building blocks that we should glue are trinions. 

These are a flavored generalization of the model in Figure \ref{E0_for_g>1} and a higher rank generalization of the genus 2 model of Figure \ref{fig:Genus2dual}. They have $2(g-1)$ $SU(3)$ gauge groups, $3(g-1)$ $SU(3)\times SU(3)$ bifundamentals and $2(g-1)$ $SU(3)\times SU(N_f+1)$ bifundamentals. The R-charges of the chiral fields under the $U(1)_R$ Cartan of the 5d $SU(2)_R$ R-symmetry are all $\frac{2}{3}$ and so their fermionic components have R-charge $-\frac{1}{3}$. Due to the fractional R-charge, we shall rescale the $U(1)_R$ R-charges by a factor of 3 so that the new R-charges are integrally quantized.\footnote{The issue here, unlike the previous case, is that a $\mathbb{Z}_3$ subgroup of the R-symmetry is actually identified with part of the center of the gauge groups. This can be seen, for instance, from the index which contains only integer R-charges. This suggests that we can still couple to background $U(1)_R$ bundles of unit magnetic charge if we compensate this with fractional $SU(3)$ magnetic monopoles. Nevertheless, we shall not consider this here, rather we opt to look at the case of coupling to a background $U(1)_R$ monopole of charge 3. This does not require the activation of fractional $SU(3)$ magnetic monopoles, and as it is still odd, should suffice in order to identify the presence of the parity anomaly.} Indeed, in this way the R-charge of the fermions in the chirals, which is the minimal charge, is $-1$, and additionally the fermion in the vector multiplet gets R-charge $3$. We can next compute
\bea
\mathrm{Tr}(U(1)^2_R)_{\text{genus }g} &=& 2(g-1)\times 8\times (3)^2+3(g-1)\times 9 + 2(g-1)\times 3(N_f+1) \nn\\
&=& 3(g-1)(2N_f+59) , 
\eea
which is indeed odd when $g-1$ is odd.

Similarly, we can take the genus $g$ model made from $\Phi$-gluing. This is obtained by gluing together $2(g-1)$ copies of the trinion theory of Figure \ref{fig:3dtrinion} by gauging $3(g-1)$ pairs of $SU(2)$ punctures and adding extra $SU(2)\times SU(N_f)$ bifundamental chirals for each $SU(2)$ gauge group of $U(1)_R$ R-charge (before the rescaling) 1. As in the previous examples, these latter chirals do not contribute to the computation of the anomaly, since their fermion components have zero R-charge, also after the rescaling. Hence, they consist of $2(g-1)$ $SU(3)$ gauge groups, $3(g-1)$ $SU(2)$ gauge nodes, $2(g-1)$ $SU(3)\times SU(N_f-2)$ bifundamentals of R-charge $\frac{2}{3}$, $6(g-1)$ $SU(3)\times SU(2)$ bifundamentals of R-charge $\frac{1}{3}$, plus the chirals added in the gluing that do not contribute. After properly rescaling the R-charges by a factor of 3 as before, we get
\bea
\mathrm{Tr}(U(1)^2_R)_{\text{genus }g} & = & 2(g-1)(8\times (3)^2 + 3(N_f-2)\times(-1)^2+3\times 3\times 2\times (-2)^2)\nonumber \\  & + & 3(g-1)\times 3\times(3)^2 = 3(g-1)(2N_f+119) . 
\eea
In all cases, we see that the anomaly coefficient agrees with the 5d expectations.

\section{Discussion}
\label{sec:Disc}

In this paper we studied 't Hooft anomaly matching between 5d SCFTs and 3d theories obtained from them by compactifying on a Riemann surface. Even though there are no anomalies for continuous symmetries in these dimensions, there are nontrivial anomalies involving discrete groups. These can be either 0-form or 1-form symmetries, and once such an anomaly is identified both in the 5d theory and in the 3d one, one can investigate the relation between them by reducing the 5d anomaly theory on the Riemann surface used in constructing the 3d theory. In particular, if the 3d theory is indeed the product of compactifying the 5d SCFT on the surface, we expect to get the 3d anomalies from the 5d ones following this procedure. Such an analysis can in principle serve three different purposes. First, it can direct us towards previously-unidentified 3d anomalies that follow from familiar 5d ones. Second, in the opposite direction, it can uncover new 5d anomalies which are required to be present in order to obtain the observed 3d anomalies upon compactification. Finally, such an analysis serves as a highly-nontrivial test for proposals relating specific 5d and 3d theories by compactification, which proved to be extremely useful in the study of compactifications from 6d to 4d (where there are also anomalies for continuous symmetries).

Focusing on the 5d and 3d theories investigated in the previous work \cite{Sacchi:2021afk,Sacchi:2021wvg,Sacchi:2023rtp}, we began with the analysis of discrete anomalies between 1-form and 0-form symmetries. We started from the case of the 5d $E_1$ SCFT, which has a known anomaly between its $\mathbb{Z}_2^{[1]}$ 1-form and $SO(3)$ 0-form symmetries. We showed how this anomaly yields the one we identified in the 3d theory obtained by a torus compactification with flux. This example therefore illustrates how given 5d and 3d anomalies can be matched by carefully integrating the 5d anomaly over a Riemann surface, and serves as a new test for the proposed relation between the corresponding theories given by compactification. 

We then turned to another interesting case in which the 5d and 3d theories have a 1-form symmetry which has a nontrivial anomaly with a 0-form symmetry. This is the case of the 5d $E_0$ SCFT, which has a $\mathbb{Z}_3^{[1]}$ 1-form symmetry and an $SU(2)_R$ 0-form symmetry. Studying the corresponding 3d theory obtained by compactifying it on a genus $g>1$ surface with no flux (found in \cite{Sacchi:2023rtp}), we found an anomaly between its $U(1)_R$ symmetry and $\mathbb{Z}_3^{[1]}$ 1-form symmetry, which implies a similar anomaly in the 5d theory. Since this 5d anomaly was not identified before, we have a new prediction for the 5d SCFT based on our compactification analysis to 3d. 

In addition to discrete anomalies between 1-form and 0-form symmetries, we also investigated anomalies involving discrete aspects of continuous 0-form symmetries. In these cases the anomalies are readily identifiable in the 3d models but are not familiar in the corresponding 5d SCFTs, and as a result yield new predictions in 5d. In particular, based on the anomalies of the 3d models obtained by compactifying on a Riemann surface we showed that the 5d $E_6$ SCFT is expected to have an anomaly between its $E_6/\mathbb{Z}_3$ symmetry and its R-symmetry, while the 5d $E_3$ SCFT is similarly expected to have an anomaly between its $PSU(3)$ symmetry and the R-symmetry. 

Finally, we considered the Witten anomaly of the 5d $SU(2)_R$ R-symmetry and investigated its fate upon reduction to 3d. We first showed that when compactifying on a genus-$g$ Riemann surface this 5d anomaly reduces to the parity anomaly of the $U(1)_R$ symmetry in 3d (where $U(1)_R$ is the 3d R-symmetry which is the Cartan of the 5d $SU(2)_R$), and then tested it in several examples. We showed how these 5d and 3d anomalies match for the torus compactification of the 5d rank-$1$ Seiberg $E_{N_f+1}$ SCFTs studied in \cite{Sacchi:2021afk}, as well as for their higher-rank generalizations discussed in \cite{Sacchi:2021wvg}. We then examined the compactification of the rank-$1$ $E_{N_f+1}$ SCFTs on higher genus surfaces, as described in \cite{Sacchi:2023rtp}, and showed how the anomaly matching works in these cases as well. 

There are several directions which will be of interest for further exploration in a future work. First, it will be interesting to extend the analysis presented in this paper to more 5d SCFTs and their corresponding 3d theories obtained by compactification, as well as to other possible anomalies. As we have shown in a number of examples, our analysis allowed us to find various new anomalies for the 5d SCFTs and so it would be interesting to study more theories with this approach. In addition, it would be interesting to investigate and test in examples the reduction of discrete anomalies for theories in other dimensions, and on compactification manifolds of possibly different dimensions. Let us also mention that we considered in this paper only standard invertible symmetries, and it would be interesting to understand how this discussion works for anomalies of higher-groups or of more exotic non-invertible symmetries (see e.g. \cite{Kaidi:2023maf}). In fact exploring the relation under dimensional reduction in these more exotic structures would also be quite interesting, and should be related to the content of this paper as such structures can in many cases be converted, usually by discrete gauging, to standard symmetries with mixed anomalies (see e.g.~\cite{Tachikawa:2017gyf}).

Finally, we found in some cases that the existence of an anomaly in the 3d theory implies a new related anomaly of the 5d SCFT, resulting in a nontrivial prediction. It would therefore be interesting to investigate whether these new anomalies can be derived directly in 5d, thereby validating these predictions. It would be interesting if their presence can be used to infer new restrictions on RG flows of 5d SCFTs. Recent years have seen some interest in the study of SUSY-breaking deformations of 5d SCFTs, see for instance \cite{BenettiGenolini:2019zth,Bertolini:2021cew,Bertolini:2022osy}, and an interesting question is if a more refined understanding of anomalies of 5d SCFTs can be put to use there.

\section*{Acknowledgments}
We would like to thank Yichul Choi, Zohar Komargodski and Shlomo Razamat for useful discussions. MS is partially supported by the ERC Consolidator Grant \#864828 “Algebraic Foundations of Supersymmetric Quantum Field Theory (SCFTAlg)” and by the Simons Collaboration for the Nonperturbative Bootstrap under grant \#494786 from the Simons Foundation. GZ is also partially supported by the Simons Foundation grant 815892. OS is supported by the Mani L. Bhaumik Institute for Theoretical Physics at UCLA. 

\bibliographystyle{JHEP}
\bibliography{ref}

\end{document}